\documentclass[fleqn,usenatbib]{mnras}

\usepackage[T1]{fontenc}
\usepackage{ae,aecompl}
\usepackage[normalem]{ulem}

\usepackage{graphicx}	
\usepackage{amsmath}	
\usepackage{amssymb}	
\usepackage{etoolbox}
\newcommand{\code}[1]{\texttt{#1}}

\newcommand{\mppnp}{\code{mppnp}} 
\newcommand{\ppn}{\code{ppn}}     

\newcommand{\spr}{\mbox{$s$-process}}
\newcommand{\sprn}{\mbox{$s$ process}}
\newcommand{\ipr}{\mbox{$i$-process}}
\newcommand{\iprn}{\mbox{$i$ process}}

\newcommand{\rprn}{\mbox{$r$ process}}

\newcommand{\numberspace}{\ensuremath{\;}}

\newcommand{\unitstyle}[1]{\ensuremath{\mathrm{#1}}}

\newcommand{\Giga}{\unitstyle{G}}

\newcommand{\Msun}{\ensuremath{\unitstyle{M}_\odot}}

\newcommand{\Gyr}{\Giga\yr}

\newcommand{\minute}{\unitstyle{min}} 
\newcommand{\hour}{\unitstyle{hr}} 
\newcommand{\yr}{\unitstyle{yr}}        

\newcommand{\unit}[2]{\ensuremath{#1\numberspace\mathrm{#2}}}

\input{vectors}
\usepackage[usenames,dvipsnames]{color}
\usepackage[]{color}

\newcommand{\arx}[1]{arXiv:1809.03666v1 [astro-ph.SR]}
\newcommand{\jphg}[1]{Journal of Physics G}

\title[The impact of reaction uncertainties on Ba to W]{The impact of (n,$\gamma$) reaction rate uncertainties of
unstable isotopes on the \ipr\ nucleosynthesis of the elements from Ba to W}

\author[P. A. Denissenkov et al.]{Pavel A. Denissenkov,$^{1,2,3\dagger}$\thanks{E-mail: pavelden@uvic.ca}
Falk Herwig$^{1,2,\dagger}$, Georgios Perdikakis$^{2,4,5}$ and
\newauthor
Hendrik Schatz$^{2,5,6,\dagger}$
\\
\\
$^{1}$Department of Physics \& Astronomy, University of Victoria, Victoria, B.C., V8W~2Y2, Canada\\
$^{2}$Joint Institute for Nuclear Astrophysics, Center for the Evolution of the Elements, Michigan State University, 640 South Shaw Lane,\\
   East Lansing, MI 48824, USA\\
$^{3}$TRIUMF, 4004 Wesbrook Mall,  Vancouver, BC, V6T~2A3, Canada\\
$^{4}$Department of Physics, Central Michigan University, Mt. Pleasant, Michigan 48859, USA\\
$^{5}$National Superconducting Cyclotron Laboratory, Michigan State University, East Lansing, MI 48824, USA\\
$^{6}$Department of Physics \& Astronomy, Michigan State University, East Lansing, Michigan 48824, USA\\
$^\dagger$NuGrid Collaboration, \href{http://nugridstars.org}{http://nugridstars.org}\\}

\date{Accepted XXX. Received YYY; in original form ZZZ}

\pubyear{2019}

\begin{document}
\label{firstpage}
\pagerange{\pageref{firstpage}--\pageref{lastpage}}
\maketitle

\begin{abstract}
The abundances of n-capture elements in the CEMP-r/s stars
agree with predictions of intermediate n-density 
nucleosynthesis, at $N_\mathrm{n}\sim 10^{13}$\,--\,$10^{15} \mathrm{cm}^{-3}$,
in rapidly-accreting white dwarfs (RAWDs). 
We have performed Monte-Carlo simulations of this \ipr\ nucleosynthesis to determine the 
impact of (n,$\gamma$) reaction rate uncertainties of 164 unstable isotopes, from $^{131}$I to
$^{189}$Hf, on the predicted abundances of 18 elements from Ba to W.
The impact study is based on two representative one-zone models with constant values of 
$N_\mathrm{n} = 3.16\times 10^{14}\ \mathrm{cm}^{-3}$ and $N_\mathrm{n} = 3.16\times 10^{13}\ \mathrm{cm}^{-3}$
and on a multi-zone model based on a realistic stellar evolution simulation of He-shell convection 
entraining H in a RAWD model with [Fe/H]\,$=-2.6$.
For each of the selected elements, we have identified up to two (n,$\gamma$) reactions having the strongest
correlations between their rate variations constrained by Hauser-Feshbach
computations and the predicted abundances, with the Pearson product-moment correlation coefficients
$|r_\mathrm{P}| > 0.15$. We find that the discrepancies between the predicted and observed
abundances of Ba and Pr in the CEMP-i star CS31062-050 are significantly diminished
if the rate of $^{137}$Cs(n,$\gamma)^{138}$Cs is reduced and the rates of $^{141}$Ba(n,$\gamma)^{142}$Ba
or $^{141}$La(n,$\gamma)^{142}$La increased. The uncertainties of temperature-dependent
$\beta$-decay rates of the same unstable isotopes have a negligible effect on the predicted abundances.
One-zone Monte-Carlo simulations can be used instead of computationally time-consuming multi-zone Monte-Carlo
simulations in reaction rate uncertainty studies if they use comparable values of $N_\mathrm{n}$.
We discuss the key challenges that RAWD simulations of \iprn\ for CEMP-i stars meet 
by contrasting them with recently published low-Z AGB \ipr. 
\end{abstract}




\section{Introduction}
\label{sec:intro}

In the \spr\ nucleosynthesis, radiative neutron captures by unstable isotopes heavier than Fe are
by definition assumed to be slower than their beta decays, so that this process proceeds alongside the n-rich boundary of
the valley of stability in the chart of nuclides. On the contrary, in the \rprn\ (n,$\gamma$) rates of
unstable isotopes are assumed to be rapid enough to exceed many beta-decay rates, in which case the ensuing nucleosynthesis involves
species far from the valley of stability towards the n-drip line \citep{burbidge:57}. All n-capture rates are proportional to
the number density of free neutrons $N_\mathrm{n}$, whose typical values for the {\it s} and {\it r} processes are
$10^7\,\mathrm{cm}^{-3}\la N_\mathrm{n}\la 10^{11}\,\mathrm{cm}^{-3}$ and $N_\mathrm{n}\ga 10^{20}\,\mathrm{cm}^{-3}$, respectively
\citep[e.g.,][]{thielemann:11,kappeler:11}.

\cite{cowan:77} suggested that some n-capture nucleosynthesis might also take place in stars at intermediate
values of $N_\mathrm{n}\sim 10^{13}$\,--\,$10^{15}\,\mathrm{cm}^{-3}$. Such an \iprn\ can occur during a He-shell flash, when
convection driven by He burning entrains H from its surrounding H-rich envelope, and neutrons are produced
in the $^{12}$C(p,$\gamma)^{13}$N(e$^+\nu)^{13}$C($\alpha$,n)$^{16}$O reaction chain, e.g. in thermally-pulsing asymptotic
giant branch (AGB) and post-AGB stars \citep{cowan:77,malaney:86,jorissen:89,Cristallo:2009cua,cristallo:16,karinkuzhi:20}.

The first strong observational evidence of an ongoing \iprn\ in a post-AGB star was the discovery of a short-term 
(on a timescale of several months) enhancement of
the surface abundances of the first peak n-capture elements Rb, Sr, Y and Zr, those with neutron numbers near magic
$N=50$, in Sakurai's object (V4334 Sagittarii) by \cite{paper:asplund1999}.
\cite{paper:herwig-2011} explained that enhancement as a result of a very late thermal pulse of a He shell that had led
to an H ingestion by He-shell convection and an \ipr\ nucleosynthesis that was interrupted by a split of the He convective zone. 
That split was found to be a result of the hydrodynamic feedback from the violent energy release in the reaction $^{12}$C(p,$\gamma)^{13}$N
operating on timescales of $\approx 15\min$ in the convective-reactive environment \citep{herwig:14}.

The first study of the impact of (n,$\gamma$) reaction rate uncertainties on the predicted abundances of the elements
with the atomic number in the range $52\leq Z\leq 63$, in many aspects similar to our studies based on one-zone Monte Carlo
simulations, results of which are presented in this paper, was done by \cite{bertolli:13}.
Later, \cite{paper:dardelet} showed that the peculiar surface abundances of the elements from Sr to Ir
found in the CEMP-r/s stars, the latter representing a sub-class of the carbon enhanced
metal-poor stars seemingly polluted by products of both {\it s} and \rprn\ \citep{bisterzo:12,abate:16}, 
could in fact be surprisingly well fitted with \ipr\ nucleosynthesis predictions.
The results of \cite{paper:dardelet} were based on one-zone
nucleosynthesis simulations in which the initial conditions, such as
H abundance and temperature had been chosen to facilitate the large
intermediate neutron density that is characteristic for the \iprn. 

The \ipr\ site creates by its nature a regime in which
  convective and nuclear burning time scales, specifically the time
  scale of the $^{12}$C(p,$\gamma)^{13}$N reaction, are equal somewhere inside
  the He-burning convection zone (e.g., see \citealt{Cristallo:2009cua} and appendix B of \citealt{paper:herwig-2011}). 
  Therefore, any physical model of the \ipr\
  conditions must incorporate both the mixing and the nuclear
  processes. One-zone models cannot accomplish this. Simple one-zone
  models aiming to explore the heavy-element nucleosynthesis at
  sufficiently high neutron density that adopt a constant temperature
  and create neutrons through H-burning and the
  $^{13}$C($\alpha$,n)$^{16}$O reactions then face a dilemma \citep{bertolli:13}. At too
  low temperature the ($\alpha$,n) reaction proceeds too slow to
  generate high $N_\mathrm{n}$ no matter how much $^{13}$C is
  available, while at too high $T$ the $^{13}$N(p,$\gamma$) reaction is
  faster than the $\beta$ decay and thereby limiting the formation of
  $^{13}$C and preventing to reach high $N_\mathrm{n}$. One simple
  trick is to just turn off the $^{13}$N(p,$\gamma$) reaction and
  thereby obtain the high neutron densities that naturally emerge in
  the more realistic multi-zone simulations, as was done in the
  one-zone models with neutron creation via charged-particle reactions
  in \cite{paper:dardelet} and \cite{mckay:20}\footnote{This network detail had not been explicitly mentioned in those two previous
  papers, although it had been eluded to in \cite{bertolli:13}. We clarify this here for the benefit of anybody who may try to
  reproduce our earlier one-zone results. This lapse in disclosure is
  otherwise however entirely inconsequential as none of the results of those
  papers relate to the details of how the neutrons are produced in the code.}. However, compared to
  multi-zone models \citep[e.g.,\ stellar-evolution based \ipr\
    models of][]{denissenkov:19} or the new 3D-based
  advective-two-stream models of \cite{Stephens:vn}, one-zone models
  cannot provide any meaningful insight into the actual formation
  process of neutrons through convective-reactive H burning, and
  therefore the better solution is to just work with a constant
  neutron density approach as done in \cite{hampel:16},
\cite{hampel:19} and also in \cite{mckay:20}. \cite{hampel:16} and
\cite{hampel:19} used a larger number of CEMP-r/s stars
and a different fitting technique to confirm the conclusions of \cite{paper:dardelet}. In all of those works
the actual stellar site of the \iprn\ remained unknown, therefore they employed a one-zone nucleosynthesis model with
a single set of specified density and temperature or a single value of $N_\mathrm{n}$ considered as free parameters.

The \ipr\ conditions are found whenever H is entrained into a
convective He-burning shell
\citep{Campbell:2010eq,paper:herwig-2011}. Such H-ingestion flashes
(HIFs) or proton-ingestion events (PIEs)\footnote{These are almost
equivalent terms used by different groups active in this research. It
is however correct that not all PIEs induce a flash. An example are
the rapidly accreting white dwarfs discussed below, in which there are
mostly PIEs but also occasionally HIFs.} may occur at various stellar evolution
phases in such objects as low-Z AGB stars
\citep[e.g.][]{Fujimoto:2000fc,Herwig:2003CNO,Iwamoto:2004ju,Campbell:2008fc,Cristallo:2009cua,Suda:2010em,cristallo:16,karinkuzhi:20},
post-AGB stars \citep[e.g.][]{iben:83a,Herwig:1999uf,MillerBertolami:2006dr,paper:herwig-2011},
super-AGB stars \citep{jones:16} or massive stars
\citep{Ritter:2018kb,Banerjee:2018jy,Clarkson:2018gq,Clarkson:2020cq}. However, 
to reach the second neutron-magic peak around Ba and also obtain the large ratio hs/ls of 
the second- (heavy \spr) to the first-peak (light \spr) n-capture elements as observed in CEMP-r/s stars 
requires to sustain the intermediate neutron density for long enough to reach high neutron exposures. 
Such conditions have not yet been self-consistently obtained in neither of these sites. 
This is mostly related to the fact that the ingested protons 
lead often to very energetic nuclear burning through reacting with $^{12}$C, and this energetic feedback will disrupt 
the He-convection zone as the H ingestion energy forms its own separate convection zone in 1D stellar evolution 
models \citep[e.g.,][]{Cristallo:2009cua,paper:herwig-2011}. 
3D stellar hydrodynamic simulations have confirmed that catastrophic instabilities can be triggered 
by the H ingestion \citep{herwig:14}. Thus, all of these \ipr\ candidates remain uncertain and 
not completely supported by simulations, especially when the goal is to reproduce the \ipr\ abundances 
in CEMP-r/s stars. This includes models of low-mass, low-metallicity AGB star HIF  stellar evolution simulations that have obtained peak neutron
densities up to $N_\mathrm{n}\sim 10^{15}\ \mathrm{cm}^{-3}$ but too small neutron exposures
\citep{Cristallo:2009cua,cristallo:16,Ritter:2018kb,karinkuzhi:20}. We will briefly discuss these models in Section~\ref{sec:agb}.

There are, as far as we know, only two exceptions. The H ingestion into
the He-core flash of a $1 \Msun$ stellar evolution model with [Fe/H]\,$ = -6.5$ yield a high enough neutron exposure, 
so that second-peak elements are produced at that low metallicity \citep{Campbell:2010eq}. However, CEMP-r/s stars that
carry the \ipr\ signature in the second-peak region have typically
higher Fe abundances ([Fe/H]\,$ > -3$, e.g. \citealt{abate:16}), and it has not been shown yet that the He-core
flash can produce second-peak \ipr\ abundances at such higher Fe abundances.

This leaves the rapidly-accreting white dwarfs \citep[RAWDs,][]{denissenkov:17,denissenkov:19} 
the only stellar evolution phase for which self-consistent stellar evolution simulations 
predict the correct conditions for second-peak \ipr\ abundances as observed in CEMP-r/s stars.
Therefore, at this time, the most likely origin of the CEMP-i stars (those CEMP-r/s stars that show a very good match 
with the \ipr\ predictions) are the RAWDs. They occur in close binary systems where a white dwarf
is accreting H-rich material at a rate $\dot{M}_\mathrm{acc}\sim 10^{-7}\,M_\odot\,\mathrm{yr}^{-1}$. The accreted H
is stably burning on the white dwarf surface, which results in an accumulation of a He shell. When its mass has reached a critical
value, the He shell experiences a thermal flash that leads to an \ipr, like in the case of Sakurai's object. The main
differences from Sakurai's object are that the H ingestion in RAWDs can last much longer (nearly a month instead of hours),
without being interrupted by a split of the He convective zone, and that, following the expansion of the accreted envelope
caused by the He-shell flash and its loss from the system either via a super-Eddington luminosity wind or through a common-envelope
interaction, the accretion resumes, and the entire cycle can repeat many times \citep{denissenkov:17}.

Unlike the \sprn, the \iprn\ involves a large number of neutron-rich isotopes  $\approx 2$ to $8$ mass numbers 
outside the valley of stability. Only theoretical predictions are available for the (n,$\gamma$) reaction rates of 
these unstable species, e.g.\ those provided by \cite{paper:rauscher} from Hauser-Feshbach model computations.
Sensitivity studies where the impact of (n,$\gamma$) rate uncertainties on the final \ipr\ abundances is determined are
critical to guide experimental efforts to better constrain (n,$\gamma$) rates on unstable nuclei \citep{larsen:19,nunes:20}.
A particular high priority are rates that have a strong impact on elemental abundances that are critical for the comparison
with observed stellar abundances.
Such studies of the impact of (n,$\gamma$) reaction rate
uncertainties of unstable isotopes on the \ipr\ nucleosynthesis have recently been presented for the first peak elements in
Sakurai's object by \cite{denissenkov:18} and for the elements with $32 \leq Z \leq 48$ in the metal-poor star HD94028 by \cite{mckay:20}
(hereafter, Paper I and Paper II).

In the present work we extend those studies to \ipr\ nucleosynthesis of the elements with $56\leq Z\leq 74$,
from Ba to W (hereafter referred to as the selected range of elements), using their abundances
in some of the CEMP-i stars to constrain the relevant neutron densities.
In Section \ref{sec:methods}, we briefly describe the analysis steps and methods developed in Papers I and II
that help to identify both the \ipr\ physical conditions and the unstable isotopes whose (n,$\gamma$) reaction rate uncertainties
have the strongest impact on the predicted abundances of heavy elements at these conditions. In order to put the results of this impact study in context we discuss in 
Section~\ref{sec:agb} the features of low-mass, low-$Z$ AGB stellar models as an alternative site of the
\ipr\ nucleosynthesis and their challenges to account for the abundances observed in CEMP-i stars. Sections \ref{sec:results} and \ref{sec:summary} summarize our results and conclusions.

\section{Analysis steps and methods}
\label{sec:methods}

\subsection{Choice of neutron densities and integration time-steps for benchmark models}

We have selected two CEMP-i stars for this study, CS31062-050 and HE2148-1247. The first star has the observed abundances of the elements
from Ba to Ir \citep{aoki:02,johnson:04} that were successfully reproduced by \cite{paper:dardelet} with a one-zone
model of \ipr\ nucleosynthesis for $\log_{10}\,N_\mathrm{n}(\mathrm{cm}^{-3}) \approx 14.5$, while the abundance distribution of the elements from Ba to Dy
in the second star \citep{cohen:03} resembles those observed in the other two stars included in the analysis of \cite{paper:dardelet}
which they fitted with the predicted \ipr\ elemental abundances for $\log_{10}\,N_\mathrm{n}(\mathrm{cm}^{-3}) \approx 13.5$.

We begin our analysis using two one-zone nucleosynthesis models that have the same fixed density
$\rho = 10^4\,\mathrm{g\,cm}^{-3}$ and temperature $T_9\equiv T/10^9\,\mathrm{K} = 0.2$ 
in which we artificially keep neutron density at the constant values $N_\mathrm{n} = 3.16\times 10^{14}\,\mathrm{cm}^{-3}$
and $N_\mathrm{n} = 3.16\times 10^{13}\,\mathrm{cm}^{-3}$ corresponding to $\log_{10}\,N_\mathrm{n}(\mathrm{cm}^{-3}) \approx 14.5$ and
$\log_{10}\,N_\mathrm{n}(\mathrm{cm}^{-3}) \approx 13.5$, respectively.

\begin{figure}
  \centering
  \includegraphics[width=\columnwidth]{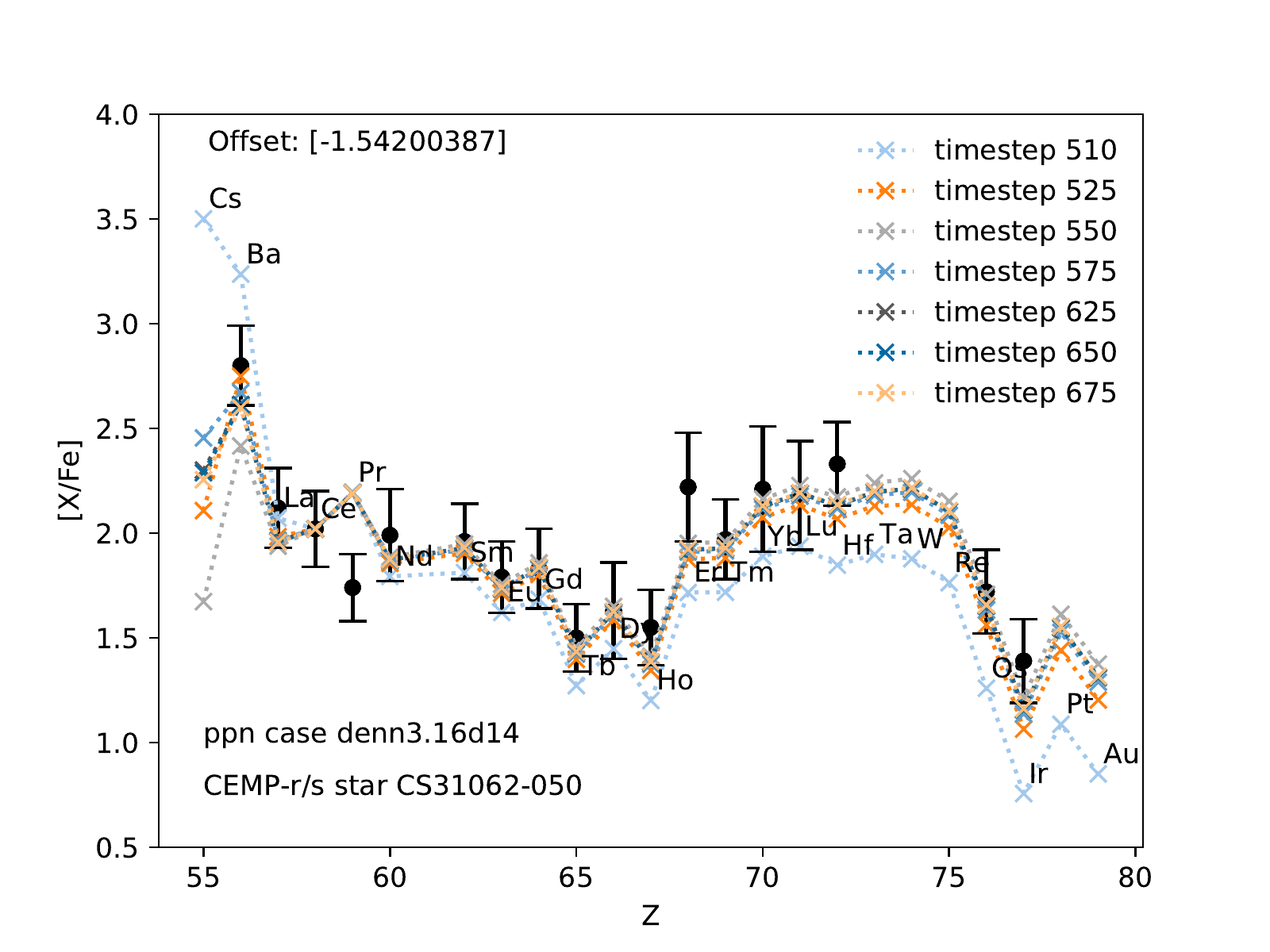}
  \caption{Selection of the minimum integration time-step for the one-zone benchmark model with $N_\mathrm{n} = 3.16\times 10^{14}\,\mathrm{cm}^{-3}$
           by comparing the predicted abundances in the selected element range  with the ones observed in the CEMP-i star CS31062-050.
           The time-step 625 (at the age of 7.5 days) is chosen because after it the predicted abundance profile stops changing, which means that
           the abundance distribution has reached a state of equilibrium characteristic for the specified neutron number density.
  }
  \label{fig:denn3p16d14cycle}
\end{figure}

\begin{figure}
  \centering
  \includegraphics[width=\columnwidth]{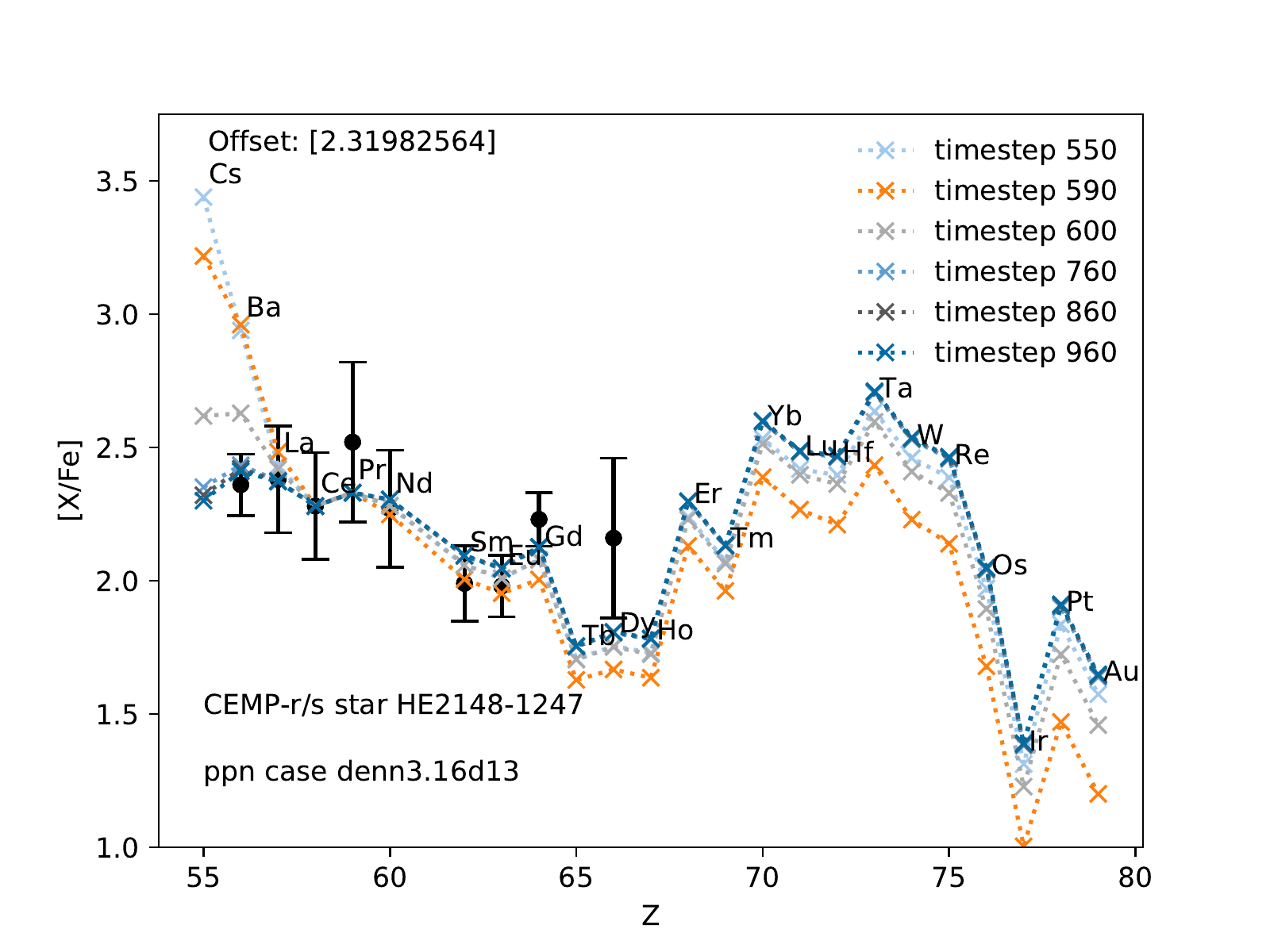}
  \caption{Same as in Figure \ref{fig:denn3p16d14cycle}, but for the one-zone benchmark model with $N_\mathrm{n} = 3.16\times 10^{13}\,\mathrm{cm}^{-3}$
           whose predicted elemental abundances are compared with those observed in the CEMP-i star HE2148-1247. In this case,
           the maximum integration time-step is chosen to be 760 (at the age of 54 days).
  }
  \label{fig:denn3p16d13cycle}
\end{figure}

Our one-zone nucleosynthesis simulations employ the same \ppn\ code from the NuGrid framework \citep{pignatari:16} 
that was used in Papers I and II. It has a small modification that allows to keep $N_\mathrm{n}$ constant. For each of the two selected stars, we call a benchmark
simulation the one that provides the best fit to its observed heavy-element abundances for the NuGrid default reaction rates 
the full list of references for which is provided in Paper I. Because these abundances are strongly enhanced compared to the Fe abundance
(black circles with error bars in Figures \ref{fig:denn3p16d14cycle} and \ref{fig:denn3p16d13cycle}), 
we can assume that their distributions have reached equilibrium states
in which abundance ratios for neighbouring elements are mostly determined by nuclear rather than stellar physics. This assumption
justifies our use of the one-zone nucleosynthesis model with a constant neutron density.

Results of reaction rate uncertainty studies are based on Monte-Carlo (MC) simulations in which selected reaction rates in the benchmark models are
randomly varied within their uncertainty ranges. To minimize the computational time of the MC simulations, 
we choose the minimum possible integration time-steps
for the one-zone benchmark models at which their predicted distributions of the heavy-element abundances have approached the assumed equilibrium
states observed in the selected stars (curves in Figures \ref{fig:denn3p16d14cycle} and \ref{fig:denn3p16d13cycle}).
The predicted abundance distributions in Figures \ref{fig:denn3p16d14cycle} and \ref{fig:denn3p16d13cycle}
have been scaled to the observed [Ce/Fe] ratio using
the pinning method from Appendix A1 of Paper II. The first step of our analysis is the adjustment of the constant neutron density and
the integration time-step for our one-zone benchmark nucleosynthesis model.

The initial chemical composition for the simulations is the \cite{paper:asplund09} solar abundance mixture scaled 
to the metallicity [Fe/H]\,$ = -2.6$, which is close to the metallicities of the two CEMP-i stars, with abundances
of $\alpha$-elements enhanced to the mean value of [$\alpha$/Fe]\,$ = +0.4$.
However, unlike the weak \iprn\ studied in Paper II, the results in this work do not depend strongly 
on the initial composition as the abundance distribution reaches equilibrium.
We suppress additional production of neutrons 
in the reactions $^{12}$C(p,$\gamma)^{13}$N(e$^+\nu)^{13}$C($\alpha$,n)$^{16}$O
by setting the initial hydrogen abundance to $X_0(\mathrm{H}) = 0$, with its mass fraction being distributed among C and O
with $X_0(\mathrm{C}) = 0.5$.

\begin{figure}
  \centering
  \includegraphics[width=0.95\columnwidth]{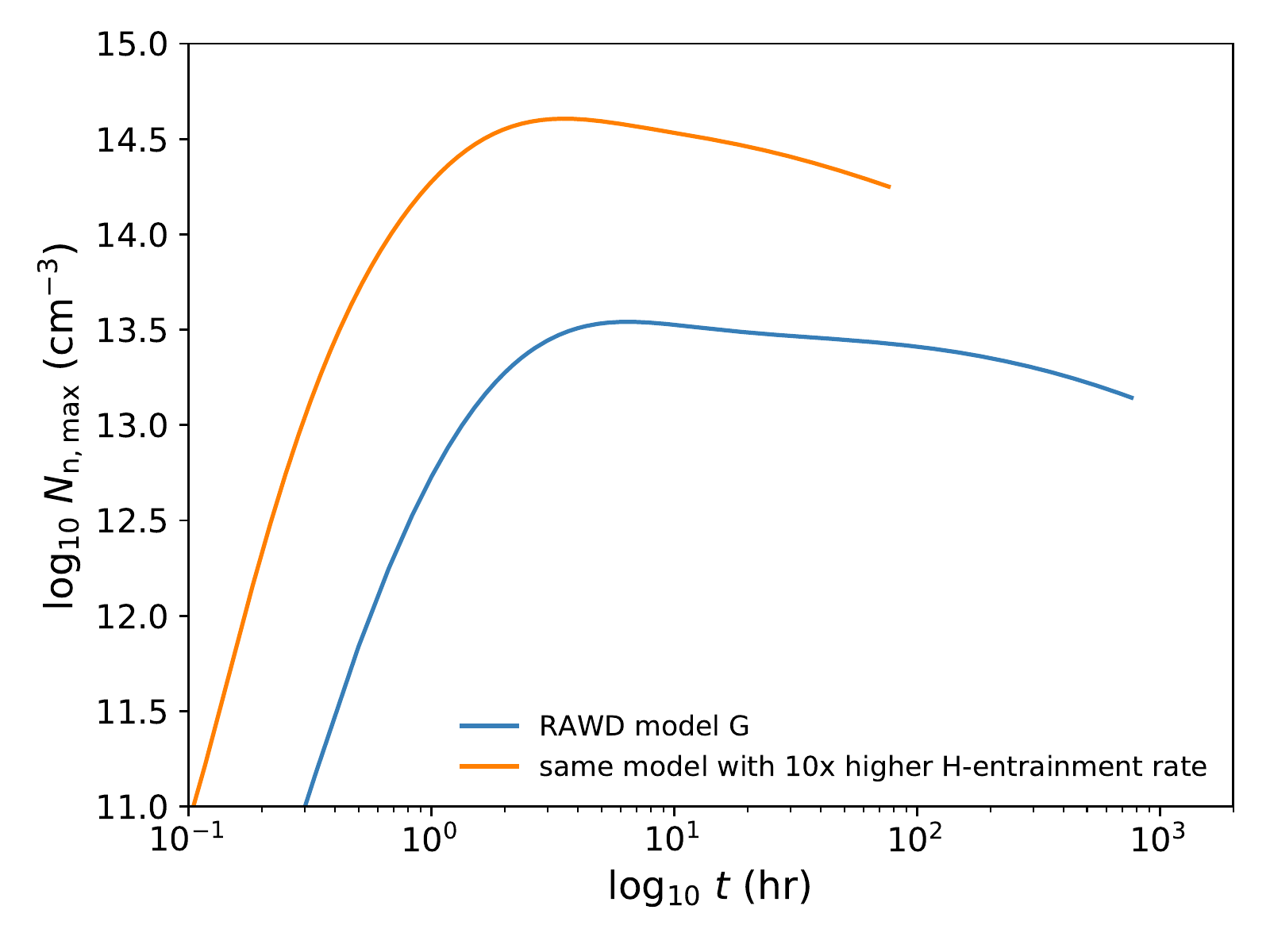}
  \caption{The evolution of the maximum neutron number density in the convective He zone of the RAWD model G (blue curve)
           from the paper of \protect\cite{denissenkov:19}. The orange curve is obtained when the H-ingestion rate has artificially
           been increased by ten times. The peaks of the two curves have the values of $N_\mathrm{n,max}$ approximately
           equal to the constant values of $N_\mathrm{n}$ used in our one-zone benchmark models from Figures \ref{fig:denn3p16d14cycle} and \ref{fig:denn3p16d13cycle}.
  }
  \label{fig:iRAWDNn}
\end{figure}

To validate the one-zone benchmark models, we have included in our analysis a multi-zone
\ipr\ nucleosynthesis model based on  
the convective He-shell structure of the RAWD model G \citep{denissenkov:19} with the same metallicity [Fe/H]\,$= -2.6$. The abundances 
computed for this model using the NuGrid code \mppnp\ \citep{pignatari:16} also fit all of
the abundances in the selected element range in the star CS31062-050 well, except Ba and Pr whose abundances are under- and over-produced, respectively 
\citep[Figure 12 in][]{denissenkov:19}. The Ba discrepancy can be caused by the fact that the maximum neutron density in
the He convective zone of the RAWD model G only reaches a value of $\log_{10}N_\mathrm{n,max}\,(\mathrm{cm}^{-3})\approx 13.5$
(blue curve in Figure \ref{fig:iRAWDNn}), which is close to the value of $N_\mathrm{n}$ in Figure \ref{fig:denn3p16d13cycle}
with a lower Ba abundance.
However, because $N_\mathrm{n,max}$ is proportional to the H-ingestion rate, a value of
$\log_{10}N_\mathrm{n,max}\,(\mathrm{cm}^{-3})\approx 14.5$, like $N_\mathrm{n}$ in Figure \ref{fig:denn3p16d14cycle}
with a higher Ba abundance, could be reached if that rate in the \ipr\ pollution source of
the star CS31062-050 were ten times higher than in the RAWD model G (orange curve in Figure \ref{fig:iRAWDNn}). 

We cannot exclude a possibility that in some RAWDs the H-ingestion rate $M_\mathrm{ing}$ is higher or lower than in the RAWD model G,
given the uncertainties of the H-ingestion rate estimates and the lack of
a detailed parameter-space study of RAWD models. Figure 5 in \cite{denissenkov:19} compares the values of $M_\mathrm{ing}$ estimated
for six 1D stellar evolution RAWD models of nearly equal masses $M_\mathrm{WD}\approx 0.73 M_\odot$, 
central temperatures $\log_{10}\,T_\mathrm{c}\approx 7.2$
and mass accretion rates $M_\mathrm{acc}\approx 1.6\times 10^{-7} M_\odot\,\mathrm{yr}^{-1}$ but different metallicities, 
$-2.6\leq \mathrm{[Fe/H]}\leq -0.7$,
with the dependence of $M_\mathrm{ing}$ on the He-shell luminosity $L_\mathrm{He}$ obtained from the 3D hydrodynamic simulations of
H ingestion that used the solar-metallicity $0.65 M_\odot$ RAWD model A from \cite{denissenkov:17} for the simulations setup. It shows that
for the values of $L_\mathrm{He}$ at the beginning of H ingestion five of the six 1D H-ingestion rates agree with
their extrapolated 3D counterparts within factors of less than $2.3$ 
and only for the RAWD model B with [Fe/H]$\,= -0.7$ the difference in 1D and 3D values of $M_\mathrm{ing}$
reaches a factor of $4.4$. From this comparison we infer that although the maximum difference 
in $M_\mathrm{ing}$ for the six RAWD models is as large as a factor of $\sim 10$ it is unlikely
to strongly depend on the initial metallicity of the WD progenitor, because neither Figure 5 nor Table 1 of \cite{denissenkov:19}
reveal any systematic dependence of $M_\mathrm{ing}$ on [Fe/H]. On the other hand, like the peak H-burning temperatures and luminosities 
in novae \citep[e.g.,][]{prialnik:95},
the He-shell luminosity and $M_\mathrm{ing}$ that is nearly proportional to $L_\mathrm{He}$ in the RAWD models may be higher for larger
$M_\mathrm{WD}$ and lower $T_\mathrm{c}$. Therefore, when considering the case of $\log_{10}N_\mathrm{n,max}\,(\mathrm{cm}^{-3})\approx 14.5$
we assume that this higher neutron density could be attained in a more massive and probably initially less luminous (colder) RAWD
than the RAWD model G.

\subsection{Selection of unstable isotopes for reaction rate uncertainty studies}

\begin{figure*}
  \centering
  \includegraphics[width=1.65\columnwidth]{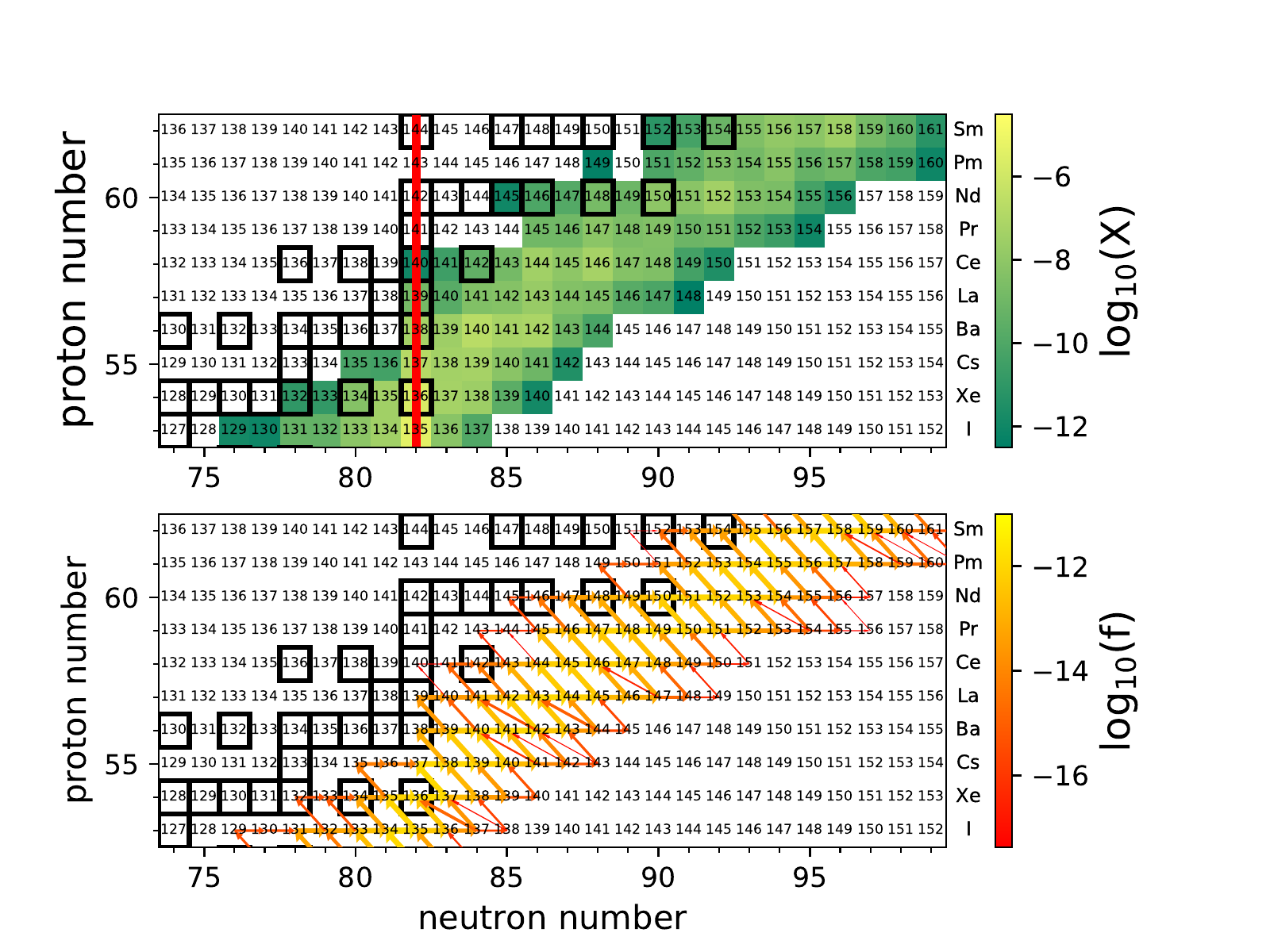}
  \caption{The mass fractions (upper panel) and reaction fluxes (lower panel) in the region of the chart of
           nuclides between I and Sm shown for the 625th integration time-step of our
           one-zone benchmark simulation of the \ipr\ with $N_\mathrm{n} = 3.16\times 10^{14}\ \mathrm{cm}^{-3}$. 
           Heavy-lined boxes show the stable isotopes, while the vertical red line marks the magic neutron number $N=82$.
             }
  \label{fig:chart1}
\end{figure*}

\begin{figure*}
  \centering
  \includegraphics[width=1.65\columnwidth]{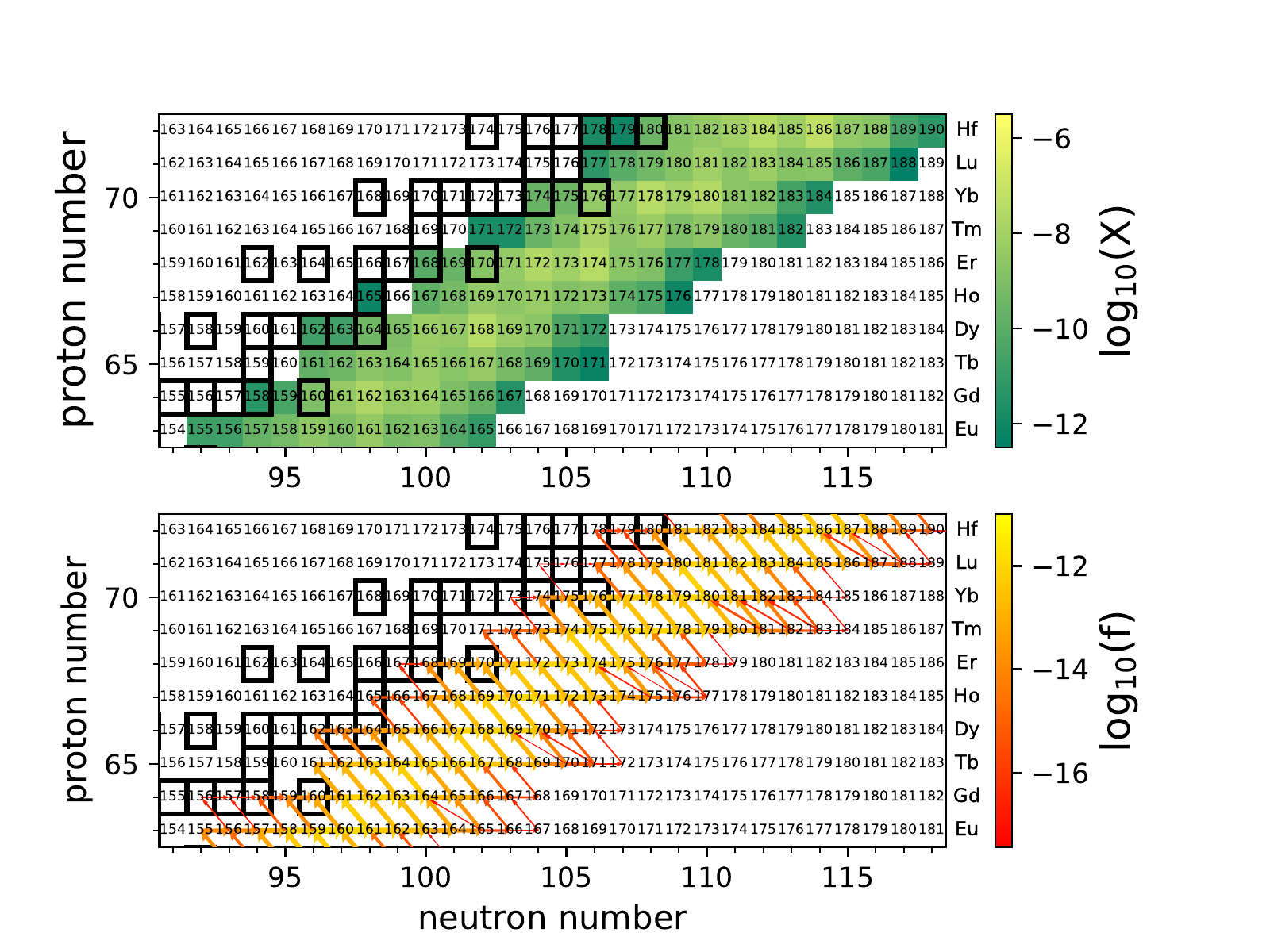}
  \caption{Same as in Fig.~\ref{fig:chart1}, but for the nuclides from Eu to Hf.
             }
  \label{fig:chart2}
\end{figure*}

At the second step, we select the unstable isotopes whose (n,$\gamma$) reaction rate variations are expected
to affect the predicted \ipr\ elemental abundances.
The mass fractions $X$ of stable and unstable isotopes and reaction fluxes $f = (\delta Y/\delta t)$, where
$Y = X/A$ for an isotope with atomic mass $A$, are shown in Figures \ref{fig:chart1} and \ref{fig:chart2} 
for the one-zone benchmark model from Figure \ref{fig:denn3p16d14cycle}. 
We use these figures to identify
164 unstable isotopes, from $^{131}$I to $^{189}$Hf, that are involved in the synthesis of
the 18 stable elements in the range from Ba to W, selected for the analysis in this work. Variations of
(n,$\gamma$) reaction and beta-decay rates of the identified unstable isotopes will change the strengths of
the reaction fluxes on the path map of the \iprn\
producing the selected elements and, as a result, their predicted abundances. 

\subsection{Maximum variation factors for (n,$\gamma$) and beta-decay rates of selected unstable isotopes}

\begin{figure}
  \centering
  \includegraphics[width=1.1\columnwidth]{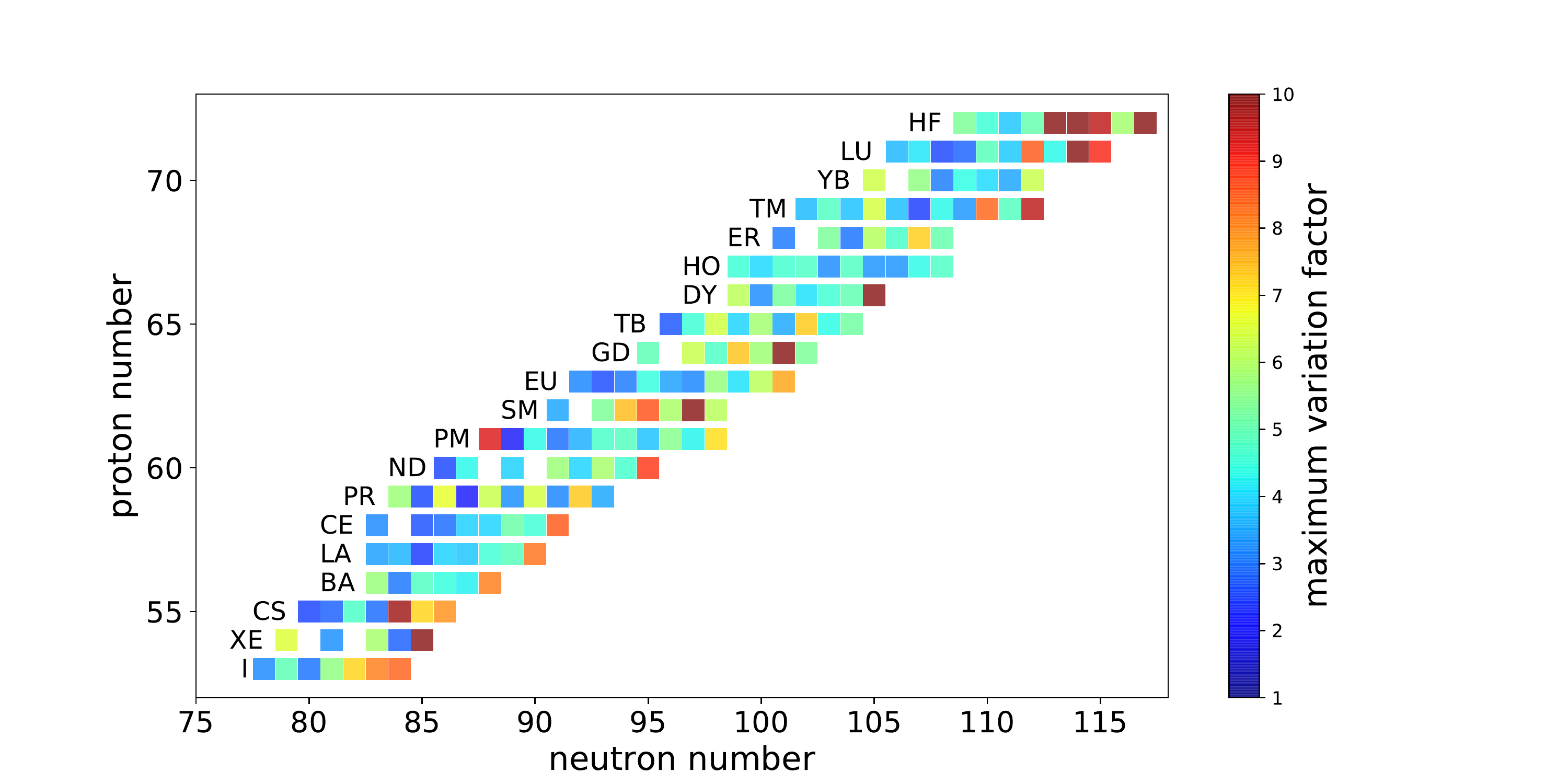}
  \caption{The 164 unstable isotopes (colour boxes) selected for the (n,$\gamma$) and beta-decay reaction rate uncertainty
           studies in this work and their corresponding maximum variation factors $v_i^\mathrm{max}$ determined 
           using Hauser-Feshbach model computations, as explained in text. 
           The following isotopes have values of $v_i^\mathrm{max}$ (given in the parentheses) exceeding the upper limit of the colour map:
           $^{139}$Xe (11), $^{159}$Sm (14), $^{165}$Gd (11), $^{171}$Dy (11), $^{185}$Lu (83), $^{185}$Hf (16), $^{186}$Hf (28), $^{189}$Hf (11).
             }
  \label{fig:george}
\end{figure}

At the third step, for each of the identified unstable isotopes a set of theoretical values of its (n,$\gamma$) rate $r_i$ is calculated with
the Hauser-Feshbach computer code \code{TALYS}\footnote{\url{http://talys.eu}} \citep{TALYS:07}. These calculations are done at a temperature of $T_9 = 0.3$
intermediate between the constant $T_9 = 0.2$ in the one-zone model and the maximum $T_9\approx 0.35$ in the multi-zone model
using 20 different combinations of the nuclear level density and $\gamma$ ray strength models from the references provided in
Table~1 of \cite{denissenkov:18}. Each of these 20 numbers represents a possible, theoretically predicted, value of
the rate, therefore we can estimate its uncertainty as $v_i^\mathrm{max} = r_i^\mathrm{max}/r_i^\mathrm{min}$,
which we define as the maximum variation factor (Figure \ref{fig:george}). 
As was shown previously \citep{liddick:16}, rate uncertainties from Hauser-Feshbach models increase
from about a factor of 2 near stability to factors of 10 a few neutrons away from stability.
The default values of $r_i$ in the \ppn\ code represent other possible
values, that are taken from \cite{paper:rauscher}.
We assume that the true value of the rate is, with equal probability, 
somewhere between its default value divided and multiplied by $v_i^\mathrm{max}$.
To study the impact of these uncertainties on the predicted abundances in the selected element range,
we perform an MC simulation with 10000 runs of the \ppn\ code in which the default rates of the 164 unstable isotopes
from Figure \ref{fig:george} are multiplied by the factors 
$f_i = (p/v_i^\mathrm{rand}) + (1-p)v_i^\mathrm{rand}$, where $p$ is assigned a value 0 or 1
with equal probability, and $v_i^\mathrm{rand}$ is randomly chosen from a uniform distribution between 1 and $v_i^\mathrm{max}$.

To investigate the impact of uncertainties of temperature-dependent beta-decay rates of the same 164 isotopes on the simulated
\ipr\ nucleosynthesis, we use Equation (7) from \cite{paper:rauscher-variation} that gives us estimates of their corresponding maximum
variation factors in the following form:
\begin{equation}
v_\beta^\mathrm{max} = \frac{^\beta U_\mathrm{g.s.}}{\tilde{G}(T)} + ^{\beta}U_\mathrm{th}\left(1-\frac{1}{\tilde{G}(T)}\right),
\end{equation}
where the ground-state and theoretical uncertainties are set to $^\beta U_\mathrm{g.s.}=1.3$, $^\beta U_\mathrm{th}=10$, and
the partition functions of the isotopes normalized to their ground state spins
$\tilde{G}(T) = G(T)/(2J_0+1))$ are taken for the temperature $T_9 = 0.3$ from Table V of \cite{paper:rauscher}.
The ground state beta decay rates of interest in this work are experimentally known and $^\beta U_\mathrm{g.s.}=1.3$ is therefore 
an overestimation of the uncertainty. However, as we show later, even these conservative uncertainties are negligible.

\subsection{Correlation analysis of results of MC simulations}

Finally, at the fourth step we do a correlation analysis of results of the MC simulation.
The MC results are $m=10000$ different sets $\left\{X_{k,j}\right\}_{j=1}^{m}$ of predicted mass
fractions $X_k$ of selected elements. Because
in the cases of interest here, unlike for Sakurai's object studied in Paper I, the \ipr\ elements had probably been accreted
by the observed stars a long time ago, we allow all unstable isotopes to decay
for \unit{1}{\Gyr} at the end of our nucleosynthesis simulations. Therefore, Tc and Pm, that do not have stable
isotopes, are not included in the analysis. The MC abundance sets differ from each other because
they are obtained using different sets of reaction rates $\left\{f_{i,j} r_i\right\}_{i=1}^{n}$, here
for the number $n=164$ of the unstable isotopes from Figure \ref{fig:george}. 
To find out which of the reaction rate variations $f_i$ has the strongest impact on the predicted
abundance $X_k$, we calculate the Pearson product-moment correlation coefficients
\begin{eqnarray}
r_\mathrm{P}(f_i,\,X_k) = \frac{\sum_{j=1}^{m}(f_{i,j}-\overline{f_i})(X_{k,j}-\overline{X_k})}
{\sqrt{\sum_{j=1}^{m}(f_{i,j}-\overline{f_i})^2}\sqrt{\sum_{j=1}^{m}(X_{k,j}-\overline{X_k})^2}},
\label{eq:rP}
\end{eqnarray}
where $\overline{f_i} = (\sum_{j=1}^{m}f_{i,j})/m$ and $\overline{X_k} = (\sum_{j=1}^{m}X_{k,j})/m$.
The benchmark simulation (the one with the NuGrid default reaction rates) is assigned the index $j=0$ and 
it uses the multiplication factors $f_{i,0} = 1$ for all reaction rates.

\section{Results}
\label{sec:results}

In this section we will first present the main results of our reaction rate uncertainty studies
based on the one-zone models from Figures \ref{fig:denn3p16d14cycle} and \ref{fig:denn3p16d13cycle}
for radiative neutron captures and beta decays (the latter only for the second model). Then we will summarize 
the results of our analysis of the multi-zone MC simulation based on the RAWD model G that reproduces the heavy-element abundance distribution
in the CEMP-i star CS31062-050 nearly as satisfactorily as the second of our one-zone models, 
i.e. without matching the observed Ba enhancement \citep[Figure 12 in][]{denissenkov:19}. 

\begin{figure}
  \centering
  \includegraphics[width=1.1\columnwidth]{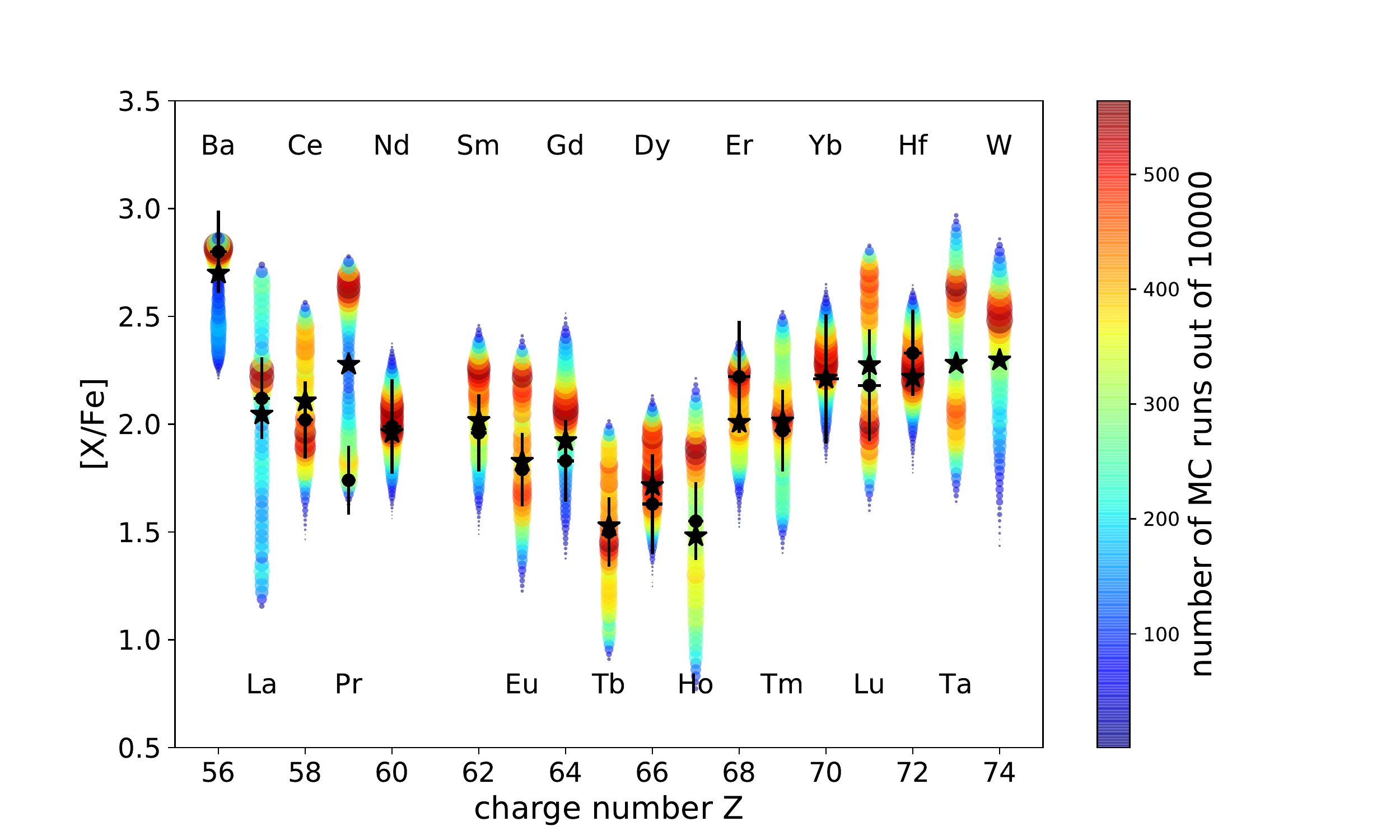}
  \caption{The circle colour- and size-coded distributions of the abundances in
           the selected element range observed in the CEMP-i star CS\,31062-050 (black circles with errorbars) predicted
           in the MC simulation based on the one-zone benchmark model with the constant $N_\mathrm{n} = 3.16\times 10^{14}\,\mathrm{cm}^{-3}$. 
           Only (n,$\gamma$) rates of the 164 unstable isotopes from Figure \ref{fig:george} have been varied.
           Black star symbols represent the abundances from the benchmark model.
           The predicted mass fractions have been mixed with the initial ones using the method described in Appendix A2 of Paper II with
           the dilution coefficient $d\approx 0.0042$ from Figure 12 in \protect\cite{denissenkov:19}.
  }
  \label{fig:uncert}
\end{figure}  

\begin{table}
        \caption{The strongest correlations between the (n,$\gamma$) reaction rate variations and
           the \ipr\ elemental abundances found in the MC simulation based on the one-zone benchmark model
           with $N_\mathrm{n}=3.16\times 10^{14}\,\mathrm{cm}^{-3}$. Only up to two correlations with
           $|r_\mathrm{P}(f_i,X_k/X_{k,0})|\geq 0.15$ are shown for which the corresponding Spearman correlation
           coefficients $r_\mathrm{S}$ are also provided.}
        \centering
        \begin{tabular}{|c|l|c|c|c|}
                \hline
                Element & Reaction & $r_\mathrm{P}$ & $r_\mathrm{S}$ & $r_\mathrm{P}(T_9=0.3)$ \\
                \hline
                Ba & {$^{135}$I} & -0.9325 & -0.9348 & -0.9248 \\
                \hline
                La & {$^{139}$Cs} & -0.6862 & -0.8500 & -0.6694 \\
                   & {$^{139}$Ba} & -0.4407 & -0.4811 & -0.4642 \\
                \hline
                Ce & {$^{140}$Cs} & -0.2134 & -0.1977 & -0.1923 \\
                   & {$^{140}$Ba} & -0.8051 & -0.9084 & -0.8124 \\
                \hline
                Pr & {$^{141}$Ba} & -0.8670 & -0.9834 & -0.8712 \\
                \hline
                Nd & {$^{144}$Ce} & -0.4964 & -0.5267 & -0.4965 \\
                   & {$^{146}$Ce} & -0.4886 & -0.5395 & -0.4779 \\
                \hline
                Sm & {$^{147}$Pr} & -0.3284 & -0.3848 & -0.3422 \\
                   & {$^{152}$Nd} & -0.7763 & -0.8493 & -0.7644 \\
                \hline
                Eu & {$^{151}$Nd} & -0.7427 & -0.8767 & -0.7464 \\
                   & {$^{153}$Nd} & -0.2122 & -0.2627 & -0.1971 \\
                \hline
                Gd & {$^{156}$Sm} & -0.5144 & -0.6305 & -0.5135 \\
                   & {$^{158}$Sm} & -0.4616 & -0.5361 & -0.4618 \\
                \hline
                Tb & {$^{159}$Sm} & -0.3931 & -0.4134 & -0.3856 \\
                   & {$^{159}$Eu} & -0.7555 & -0.8639 & -0.7561 \\
                \hline
                Dy & {$^{161}$Eu} & -0.3260 & -0.3336 & -0.3293 \\
                   & {$^{162}$Gd} & -0.7119 & -0.7978 & -0.6951 \\
                \hline
                Ho & {$^{165}$Tb} & -0.7539 & -0.9001 & -0.7359 \\
                   & {$^{165}$Dy} & -0.2322 & -0.2599 & -0.2492 \\
                \hline
                Er & {$^{167}$Tb} & -0.2244 & -0.2300 & -0.2225 \\
                   & {$^{168}$Dy} & -0.7822 & -0.8503 & -0.7771 \\
                \hline
                Tm & {$^{169}$Dy} & -0.5086 & -0.5350 & -0.5122 \\
                   & {$^{169}$Ho} & -0.6728 & -0.7867 & -0.6683 \\
                \hline
                Yb & {$^{172}$Er} & -0.5862 & -0.6386 & -0.5802 \\
                   & {$^{174}$Er} & -0.3986 & -0.4244 & -0.3879 \\
                \hline
                Lu & {$^{175}$Er} & -0.2175 & -0.2372 & 0.2210 \\
                   & {$^{175}$Tm} & -0.8166 & -0.9446 & 0.8135 \\
                \hline
                Hf & {$^{177}$Tm} & -0.3517 & -0.3649 & -0.3350 \\
                   & {$^{178}$Yb} & -0.6911 & -0.7609 & -0.6937 \\
                \hline
                Ta & {$^{181}$Lu} & -0.7691 & -0.9199 & -0.7620 \\
                   & {$^{181}$Hf} & -0.2634 & -0.2892 & 0.2670 \\
                \hline
                W  & {$^{184}$Hf} & -0.5090 & -0.6002 & -0.5075 \\
                   & {$^{186}$Hf} & -0.5041 & -0.5807 & -0.4946 \\
                \hline
        \end{tabular}
        \label{table:tab1}
\end{table}

\subsection{One-zone model with $N_\mathrm{n} = 3.16\times 10^{14}\,\mathrm{cm}^{-3}$}

For this model, we have randomly varied only (n,$\gamma$) rates.
Before calculating and analyzing correlation coefficients, we plot distributions of abundances
predicted in the MC simulation for each of the elements in the selected range of $Z$ (Figure \ref{fig:uncert}).
To make this plot we have used the dilution method from Appendix A2 of Paper II where we
demonstrated that the pinning and dilution methods, the former having been used in Figures 
\ref{fig:denn3p16d14cycle} and \ref{fig:denn3p16d13cycle}, 
can be equivalently used in this case where the overabundances of the n-capture elements in question are so large (typically $\geq 2\mathrm{dex}$). 
Most elements respond to reaction-rate variations with an asymmetric distribution, in several cases even double-peaked. 
These are signatures of their abundances being
strongly affected by variations of individual (n,$\gamma$) rates. To find out what reactions have the strongest
impact on a given elemental abundance, and therefore are likely to be responsible for its asymmetric distribution,
we display in Table \ref{table:tab1} one or two reactions with the maximum Pearson correlation
coefficients, such that $|r_\mathrm{P}(f_i,X_k/X_{k,0})| \geq 0.15$, for each of the elements in the selected range of $Z$, where
$X_{k,0}$ are their abundances predicted by the benchmark model. The lower limit $0.15$ for the correlation coefficient 
allows us to reveal most of the (n,$\gamma$) reactions responsible for asymmetric distributions of the
predicted elemental abundances. In the fourth column of Table \ref{table:tab1} we provide Spearman correlation coefficients.
The fact that for most of the selected elements the Spearman coefficients, examining the monotonicity of the found correlations, 
exceed by the absolute magnitude their corresponding Pearson coefficients, the latter estimating the linearity of the correlations,
tells us that the correlations are not linear but significant. Finally, to check that our results are not very sensitive
to the choice of the characteristic temperature at which the \ipr\ is simulated in the one-zone model, we have changed
it from $T_9 = 0.2$ to $T_9 = 0.3$, and the Pearson coefficients for the latter case are presented in the last column of Table \ref{table:tab1}.
It is seen that they insignificantly differ from the Pearson coefficients for the case of $T_9 = 0.2$.

Table I is our main result and identifies the most important neutron capture rate uncertainties affecting each of the produced elemental abundances. 
The largest correlation coefficients found for each element identify the unstable isotopes whose
reaction rate variations within their adopted uncertainties have the strongest impact on this element's predicted abundance.

\begin{figure}
  \centering
  \includegraphics[width=0.8\columnwidth]{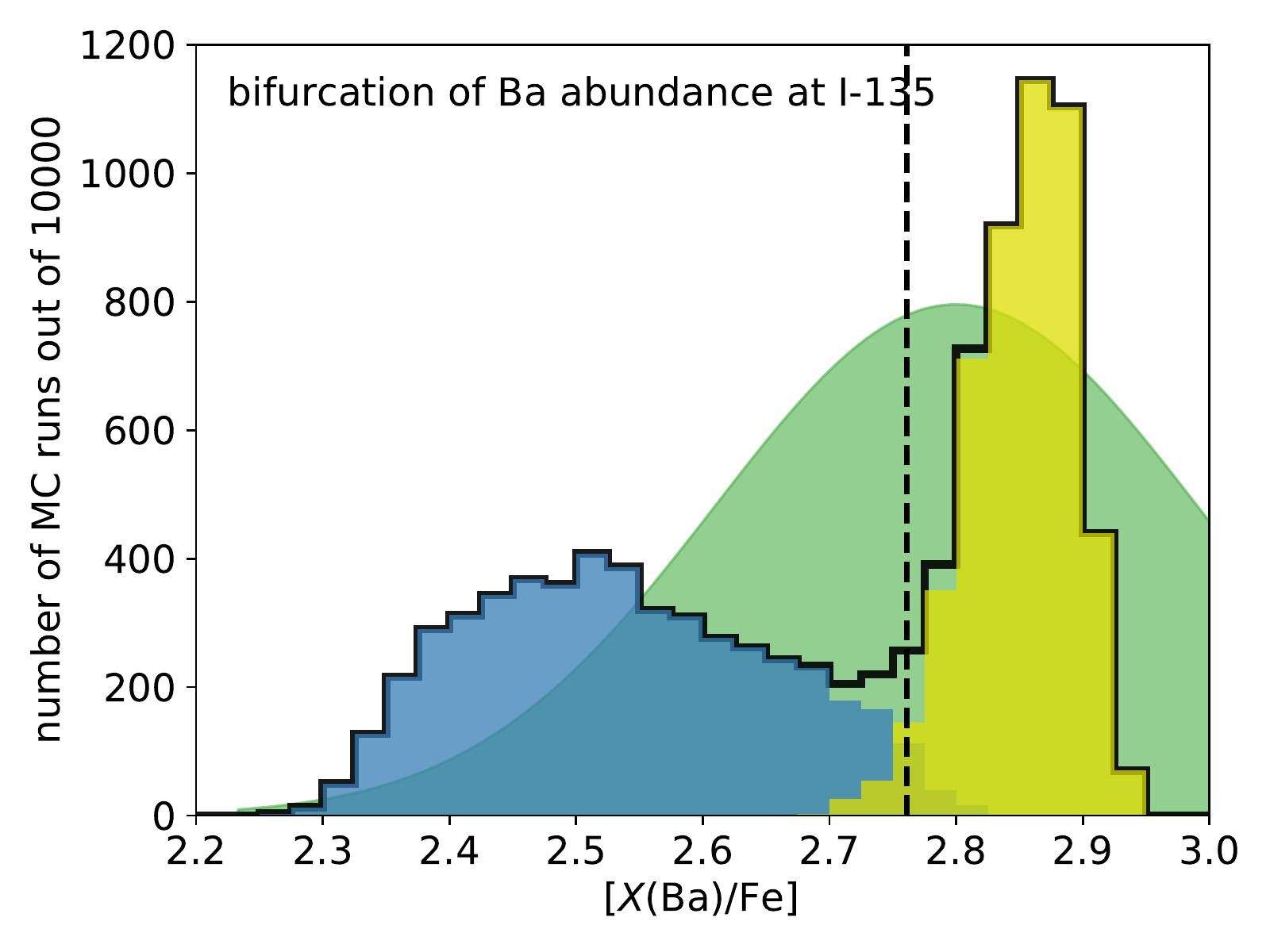}
  \includegraphics[width=0.8\columnwidth]{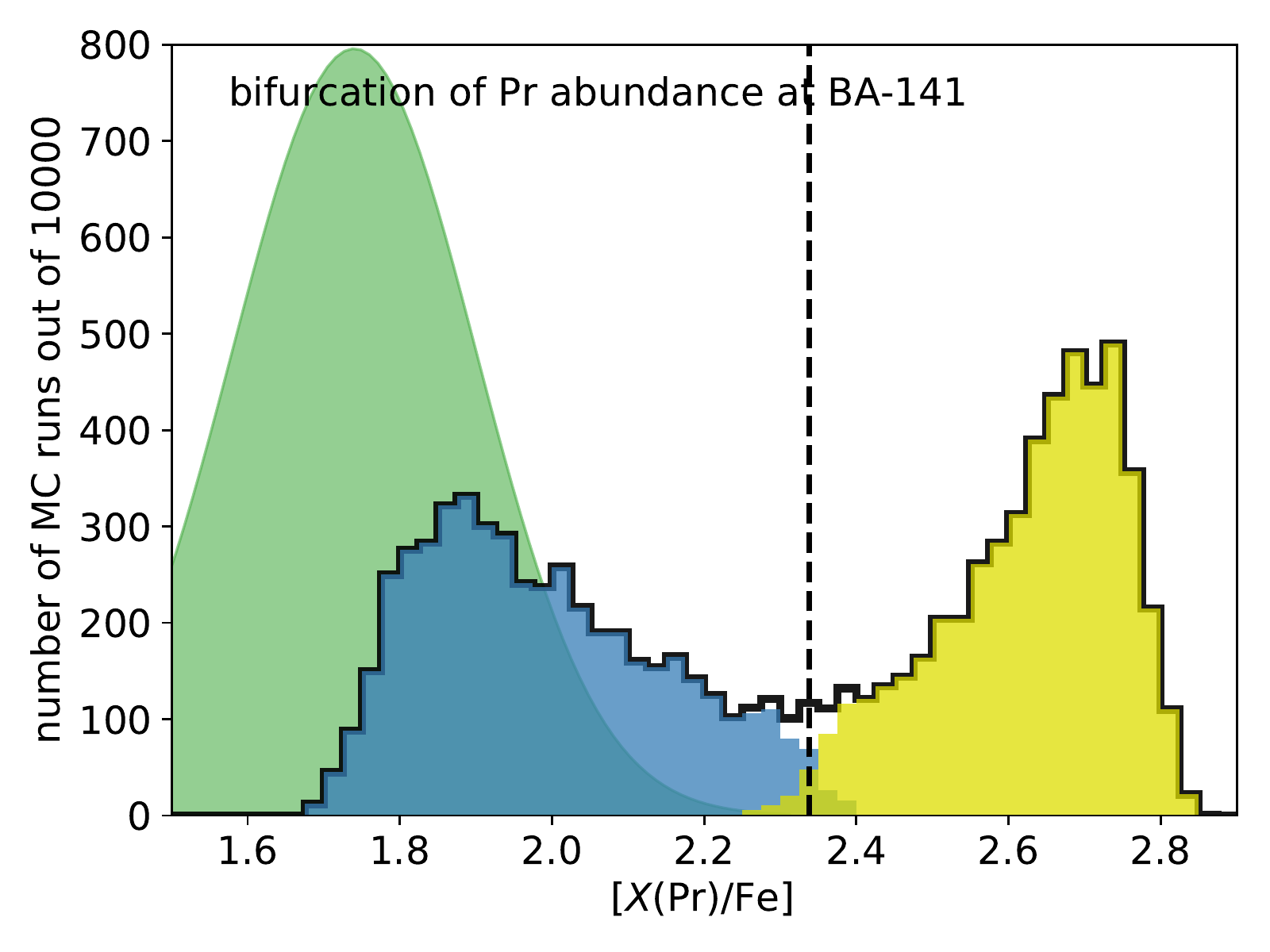}
  \caption{Comparison of the distributions of the Ba (top panel) and Pr (bottom panel) abundances obtained in the MC simulation (thick black line)
           based on the one-zone benchmark model with $N_\mathrm{n} = 3.16\times 10^{14}\,\mathrm{cm}^{-3}$ (vertical dashed line)
           with the abundances of Ba and Pr reported for the star CS31062-050 by \protect\cite{johnson:04} (green-shaded Gaussian curves).
           The single-peak distributions, blue and yellow histograms, correspond to
           rate multiplication factors $f_i > 1$ and $f_i < 1$ for the reactions
           $^{135}$I(n,$\gamma)^{136}$I (top panel) and $^{141}$Ba(n,$\gamma)^{142}$Ba (bottom panel), as identified by their strongest correlation
           coefficients in Table \ref{table:tab1} for Ba and Pr, respectively.}
  \label{fig:hist2}
\end{figure}  

\begin{figure}
  \centering
  \includegraphics[width=\columnwidth]{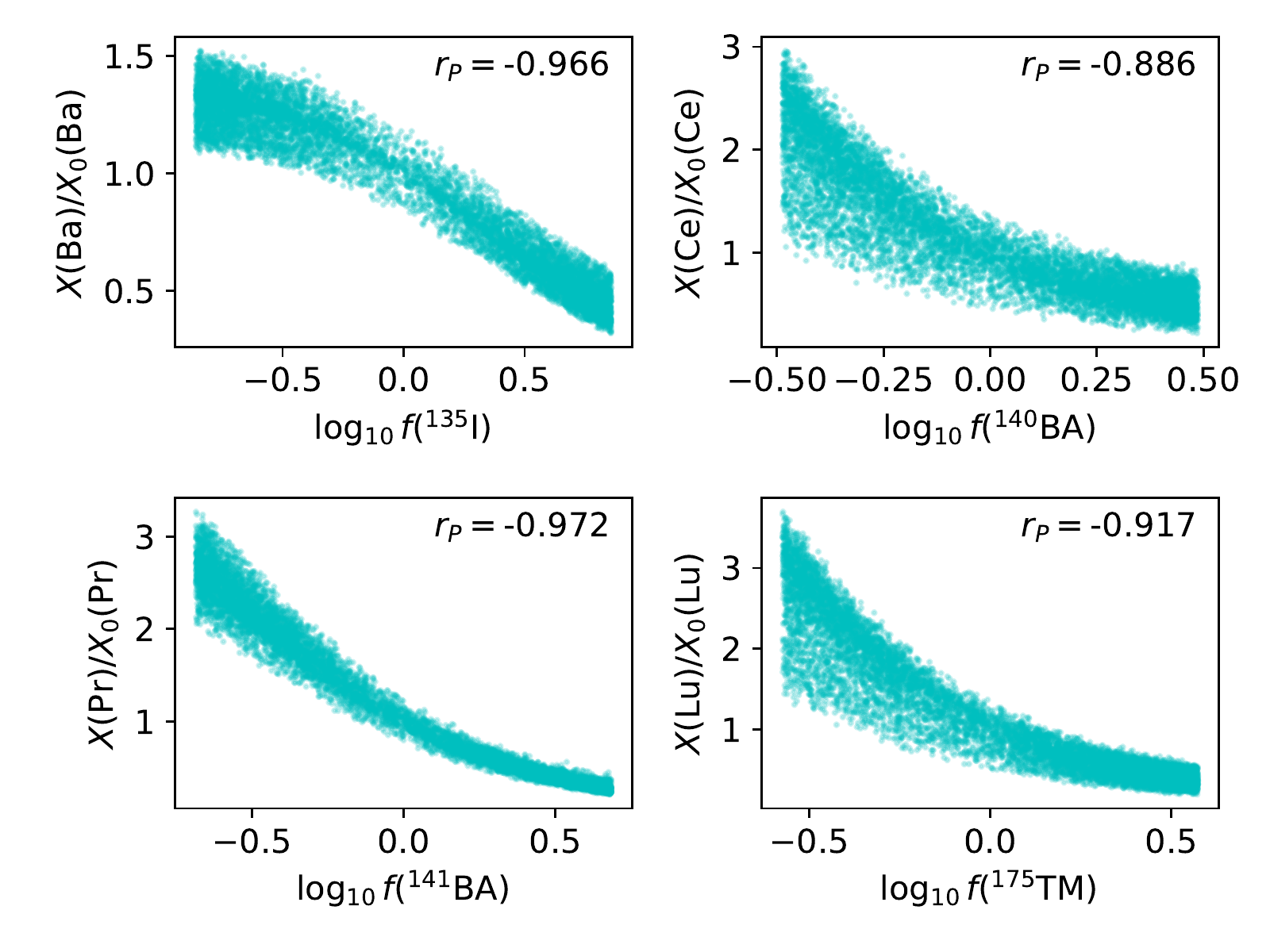}
  \caption{The distributions of the Ba, Ce, Pr, and Lu elemental abundances predicted in the one-zone MC simulations with 
           $N_\mathrm{n} = 3.16\times 10^{14} \mathrm{cm}^{-3}$ as functions of the reaction rate multiplication factors $f_i$
           with which they have the strongest correlations. The small differences between the corresponding
           correlation coefficients in this plot and in Table~\ref{table:tab1} are caused by the replacement of $f_i$ with $\log_{10}\,f_i$
           in the correlation analysis.
             }
  \label{fig:corr1}
\end{figure}

The thick black lines
in Figure \ref{fig:hist2} are the distributions of the Ba and Pr abundances obtained in the MC simulation based on
the one-zone benchmark model with $N_\mathrm{n} = 3.16\times 10^{14}\,\mathrm{cm}^{-3}$. They are compared with the abundances of Ba and Pr reported
for the star CS31062-050 by \cite{johnson:04} that are represented by the green-shaded Gaussian curves for which the mean values and standard deviations
are the same that we used in Figure \ref{fig:uncert} to plot black circles with errorbars. 
The double-peak distribution of the Ba abundances in the MC simulation can be separated into
two distinct single-peak distributions (the blue and yellow histograms)
if we divide the results of the MC simulation into two groups, the one with the multiplication factors 
$f_i > 1$ and the other with $f_i < 1$ for the rate of the reaction $^{135}$I(n,$\gamma$)$^{136}$I whose variation has the strongest correlation with
the predicted Ba abundance (Table \ref{table:tab1}, see also \citealt{bertolli:13} and \citealt{cristallo:16}). 
A similar result is obtained for Pr when we use the rate of the reaction $^{141}$Ba(n,$\gamma)^{142}$Ba identified
for it in Table \ref{table:tab1}. We see that the Ba abundance predicted with the benchmark model (the vertical dashed line) agrees very well
with the observed Ba abundance. However, this is not true for Pr. In this case the higher than the default rate of the reaction
$^{141}$Ba(n,$\gamma)^{142}$Ba could help to significantly reduce the discrepancy between its predicted and observed abundances.

Figures \ref{fig:corr1} and \ref{fig:corr2} show the distributions of the elemental abundances of Ba, Ce, Pr and Lu obtained
in our one-zone MC simulations as functions of the (n,$\gamma$) reaction rate multiplication factors $f_i$ with which they have the strongest
correlations. The Pearson product-moment correlation coefficients in the four panels of these figures show how well compared data (anti-)correlate, 
but unfortunately they do not provide quantitative estimates of the slopes of the correlations that would also be useful to have. 
A comparison of these figures with similar ones for other elemental abundances shows that distinctly visible
slopes in correlations like these appear only when $|r_\mathrm{P}|\ga 0.4$.

\begin{figure}
  \centering
  \includegraphics[width=1.1\columnwidth]{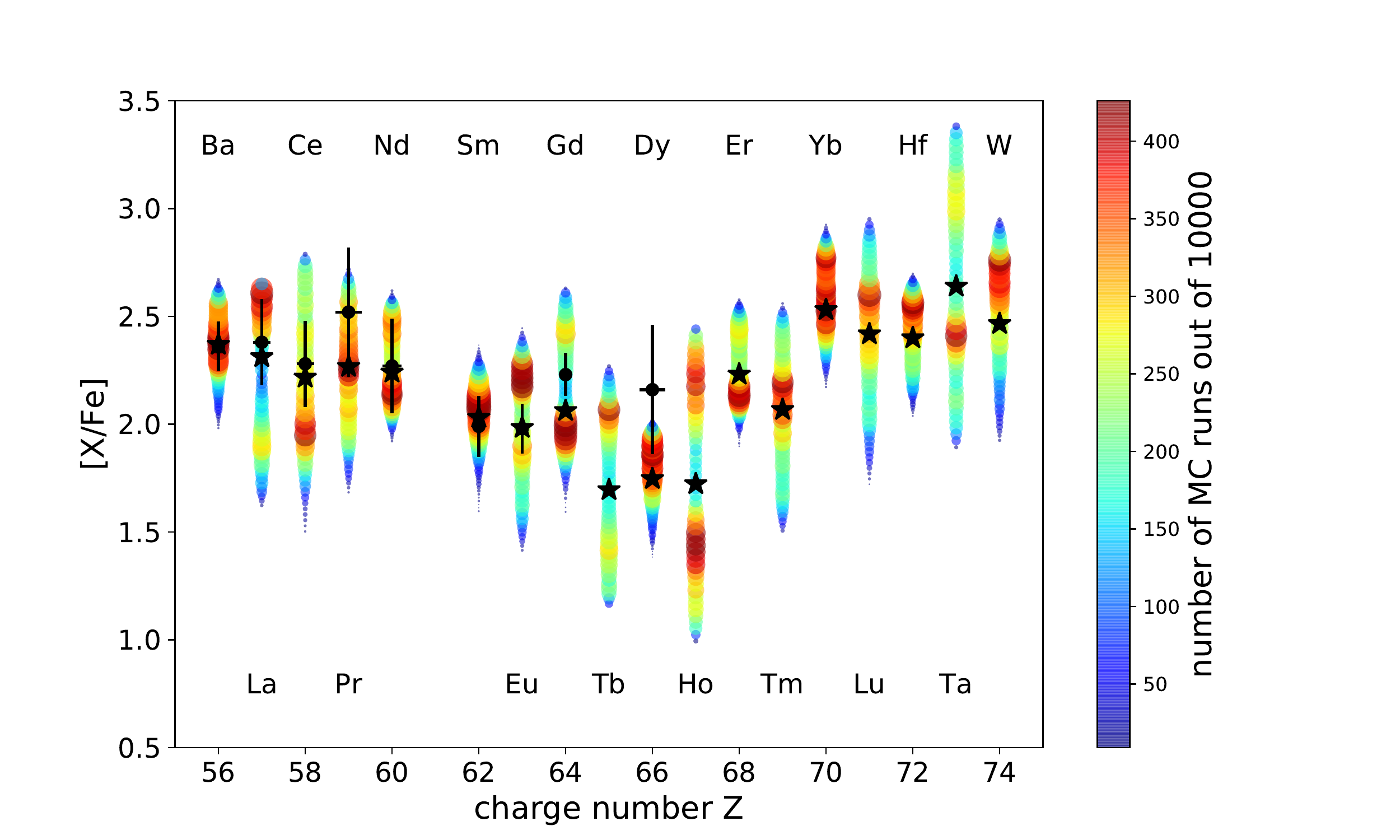}
  \caption{Same as in Figure \protect\ref{fig:uncert}, but for the one-zone benchmark model with $N_\mathrm{n} = 3.16\times 10^{13}\,\mathrm{cm}^{-3}$
           compared with the observed abundances in the CEMP-i star HE2148-1247.}
  \label{fig:uncert4}
\end{figure}  

\begin{figure}
  \centering
  \includegraphics[width=1.1\columnwidth]{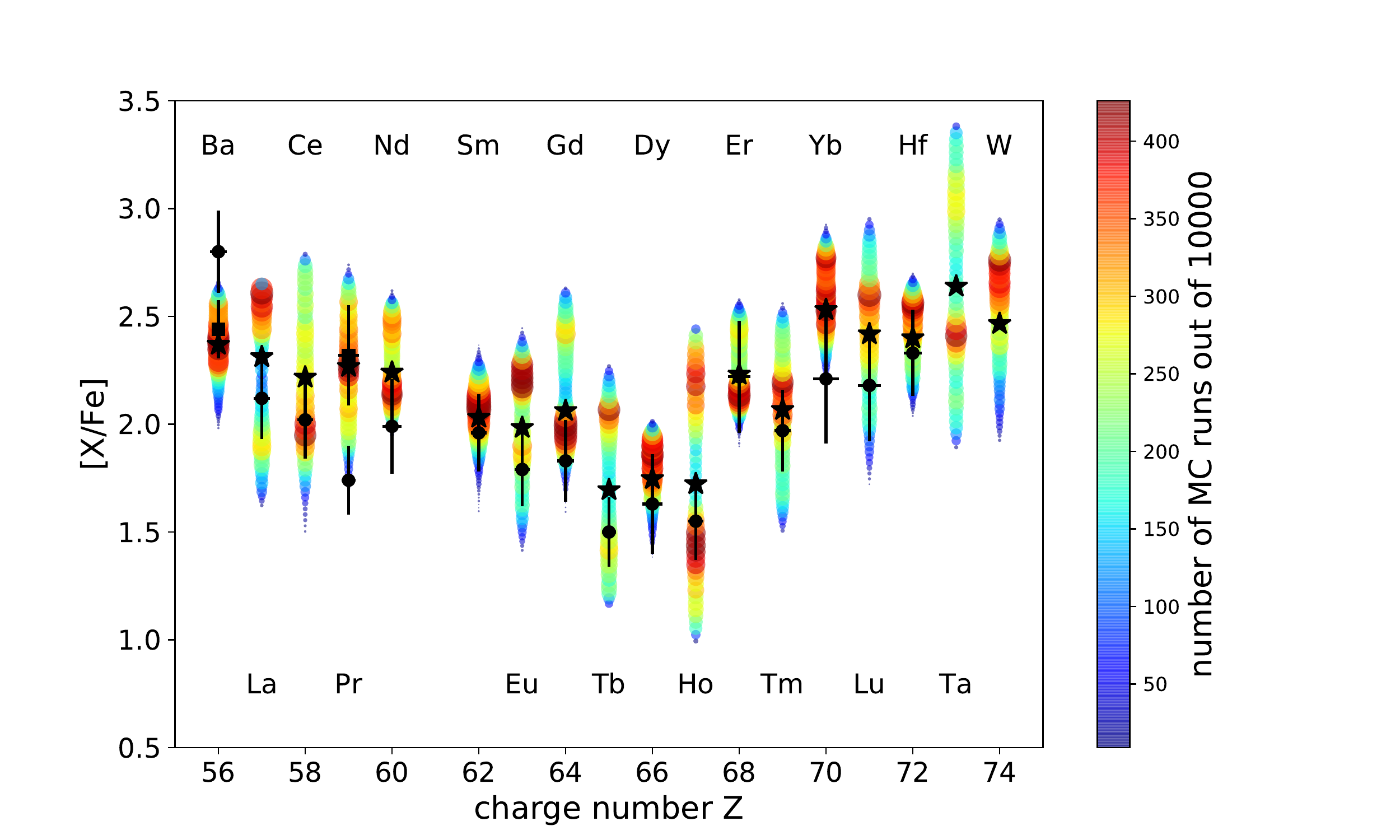}
  \caption{Same as in Figure \protect\ref{fig:uncert}, but for the one-zone benchmark model with $N_\mathrm{n} = 3.16\times 10^{13}\,\mathrm{cm}^{-3}$.
           Black squares with error bars show mean values and standard deviations for the abundances of Ba and Pr obtained in the MC simulation.}
  \label{fig:uncert2}
\end{figure}  

\begin{figure*}
  \centering
  \includegraphics[width=0.8\columnwidth]{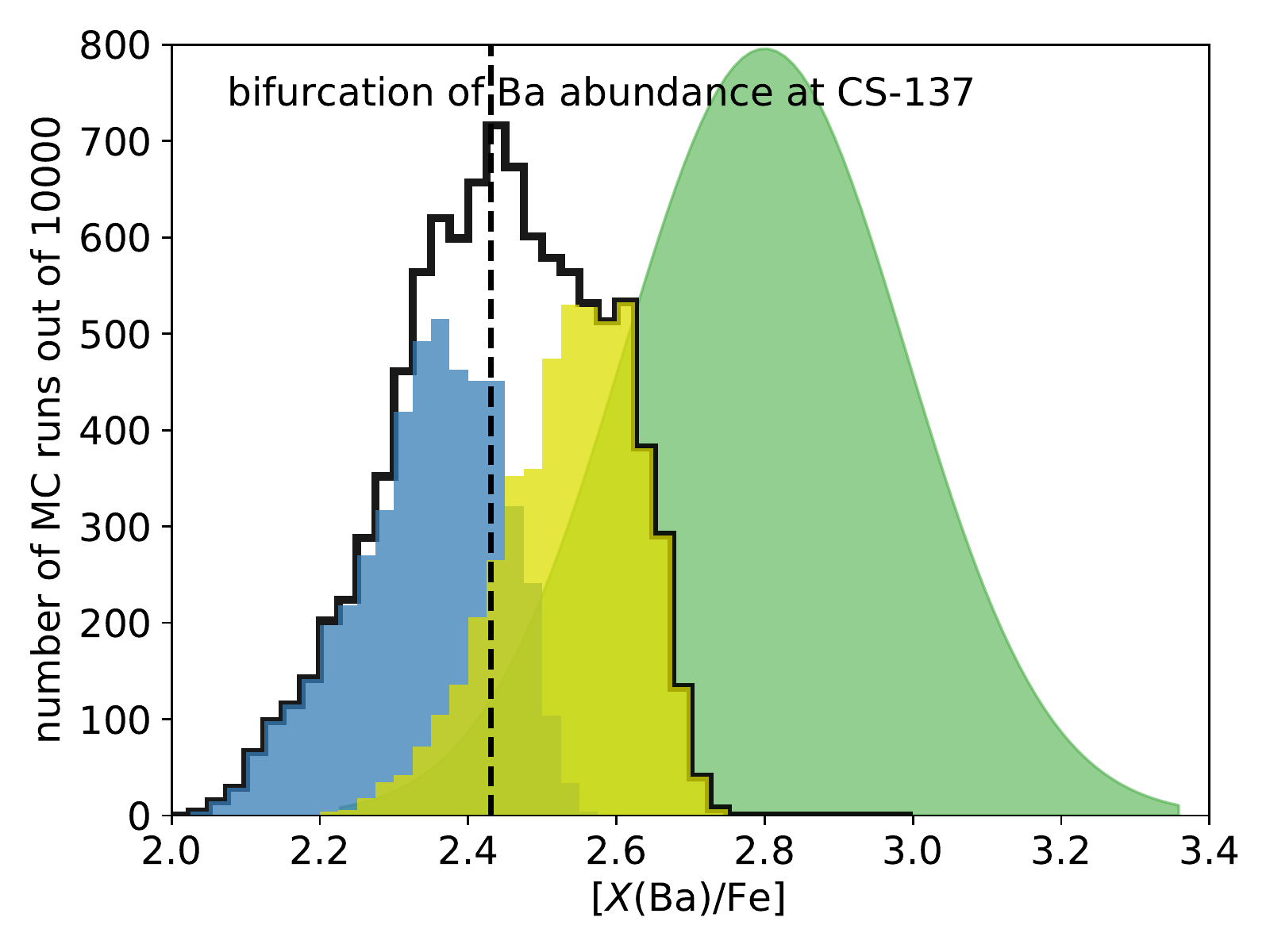}
  \includegraphics[width=0.8\columnwidth]{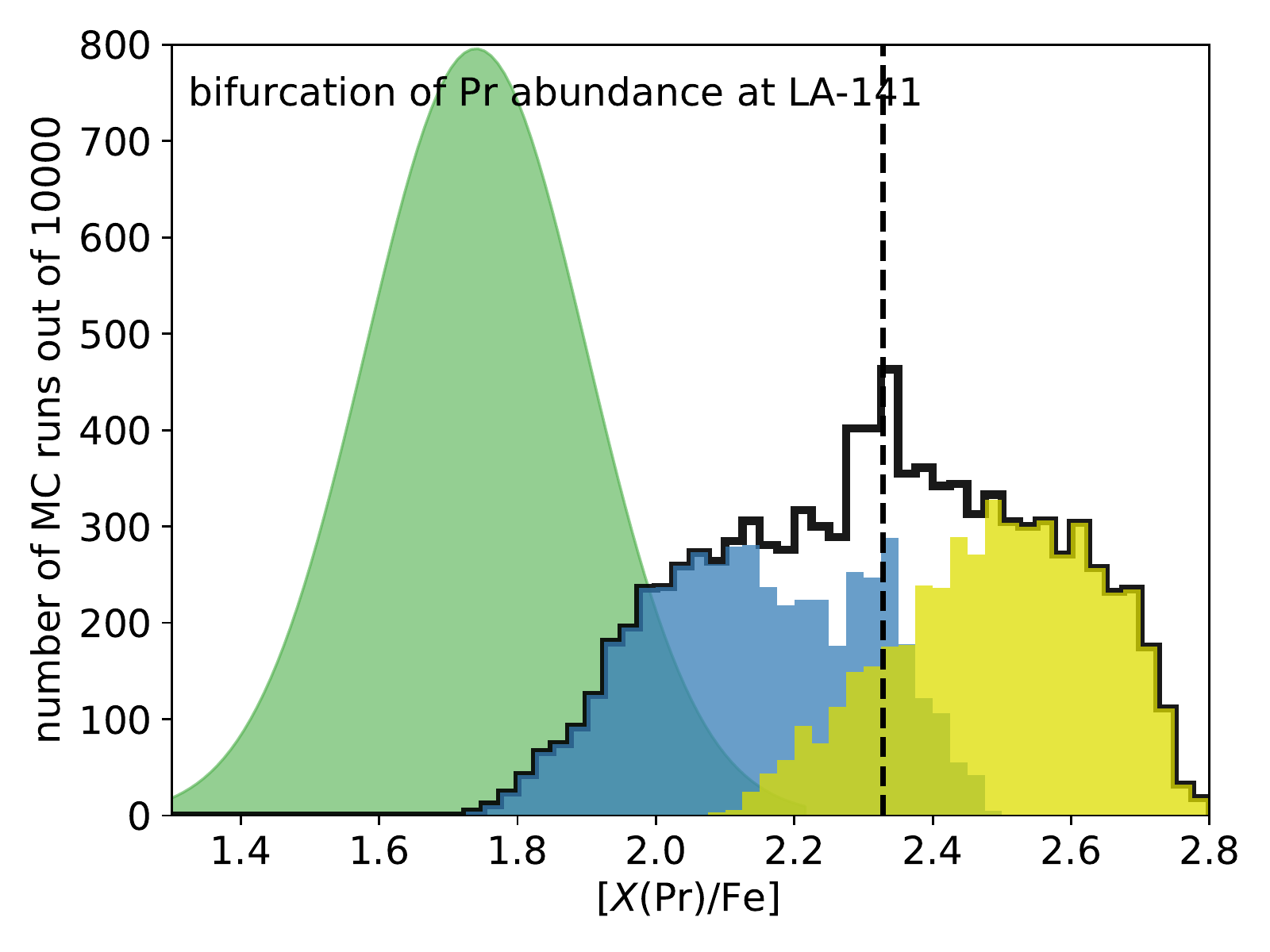}
  \caption{Comparison of the distributions of the Ba (left panel) and Pr (right panel) abundances obtained in the MC simulation (thick black lines)
           based on the one-zone benchmark model with $N_\mathrm{n} = 3.16\times 10^{13}\,\mathrm{cm}^{-3}$ (vertical dashed line)
           with the abundances of Ba and Pr reported for the star CS31062-050 by \protect\cite{johnson:04} (green-shaded Gaussian curves).
           The single-peak distributions, blue and yellow histograms, correspond to
           rate multiplication factors $f_i > 1$ and $f_i < 1$ for the reactions
           $^{137}$Cs(n,$\gamma)^{138}$Cs (left panel) and $^{141}$La(n,$\gamma)^{142}$La (right panel), 
           as identified by their strongest correlation
           coefficients in Table \ref{table:tab2} for Ba and Pr, respectively.
          }
  \label{fig:hist22}
\end{figure*}  

\begin{figure}
  \centering
  \includegraphics[width=1.1\columnwidth]{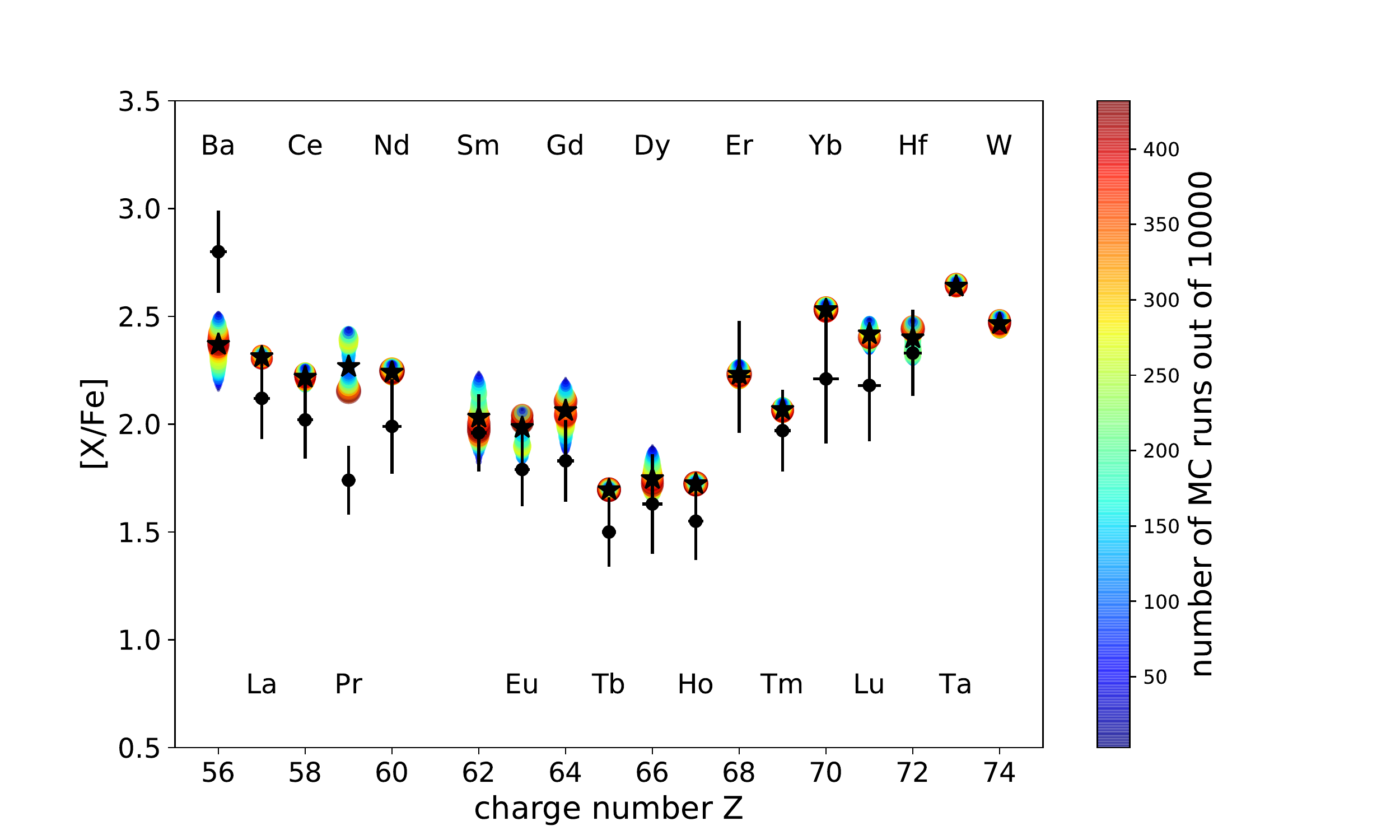}
  \caption{Same as in Figure \protect\ref{fig:uncert2}, but
           in this case only beta-decay rates of the 164 unstable isotopes from Figure \ref{fig:george} have been varied in the MC simulation.}
  \label{fig:uncert2_bdecay}
\end{figure}  

\subsection{One-zone model with $N_\mathrm{n} = 3.16\times 10^{13}\,\mathrm{cm}^{-3}$}

This benchmark model has the neutron density adjusted to reproduce the abundances of the elements in the selected range of $Z$
in the star HE2148-1247 (Figures~\ref{fig:denn3p16d13cycle} and \ref{fig:uncert4}).
However, the $N_\mathrm{n} = 3.16\times 10^{13}\,\mathrm{cm}^{-3}$ model also reproduces most of the abundances in CS31062-050 (Figure \ref{fig:uncert2}). 
The exceptions are Ba and Pr, and the Ba case is the sole reason a higher neutron density was assumed for that star. 
However, the observed Ba abundance has large systematic errors that signficantly exceed the error bars shown. 
For example, \cite{lai:07} obtain for CS31062-050 [Ba/Fe]\,$= 2.37$, significantly lower than the [Ba/Fe]\,$= 2.8$ 
from the same group published in \cite{johnson:04} that we use here. This lower [Ba/Fe] value would be in agreement 
with the prediction of our lower neutron density model. Therefore, in addition to the nuclear uncertainties, 
observational uncertainties need to be addressed to constrain the neutron density more stringently. 
With the current uncertainties, both neutron density models can be considered potential \ipr\ sources for CS31062-050.
However, the significant decrease of $N_\mathrm{n}$
between the first and the second of our one-zone \ipr\ nucleosynthesis simulations
changes the \ipr\ band path and therefore the important (n,$\gamma$) reactions.
This is reflected in Table \ref{table:tab2}
that for most elements lists different (n,$\gamma$) reactions having the strongest impact on their predicted abundances than Table \ref{table:tab1}.
For example, instead of $^{135}$I, it is now $^{137}$Cs that has the strongest impact on the Ba abundance, and because the benchmark value of [Ba/Fe]
is now much lower than the one reported for the star CS31062-050 by \cite{johnson:04}, a lower rate of the reaction $^{137}$Cs(n,$\gamma)^{138}$Cs 
would reduce this discrepancy (the yellow histogram in the left panel of Figure \ref{fig:hist22}). 
For Pr, it is now an increase of the $^{141}$La(n,$\gamma)^{142}$La reaction rate that is required to diminish the discrepancy
between its predicted and observed abundances (the blue histogram in the right panel of Figure \ref{fig:hist22}).

For this benchmark model we have also done an MC simulation in which only beta-decay rates of the 164 unstable isotopes 
from Figure \ref{fig:george}
have been varied. The results of this simulation plotted in Figure \ref{fig:uncert2_bdecay} show that these variations have a much weaker effect
on the predicted abundances of the elements in the selected range of $Z$ as compared to the variations of the (n,$\gamma$) rates for the same isotopes.
This justifies our decision to not include them into the present reaction rate uncertainty studies. 

\begin{figure}
  \centering
  \includegraphics[width=\columnwidth]{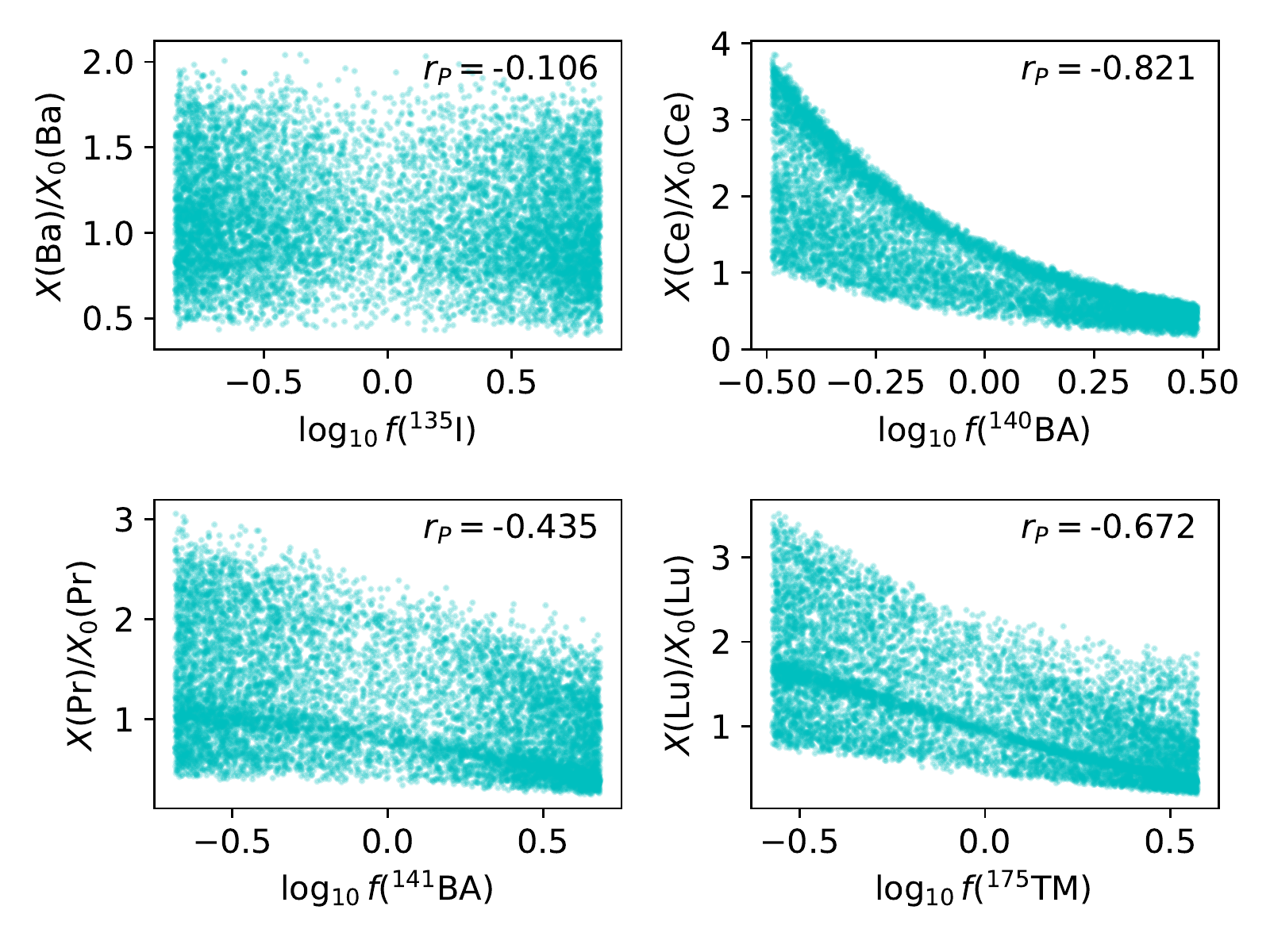}
  \caption{Same as in Figure~\ref{fig:corr1}, but for the one-zone model with
           $N_\mathrm{n} = 3.16\times 10^{13} \mathrm{cm}^{-3}$.
             }
  \label{fig:corr2}
\end{figure}

If we used the elemental abundances predicted by the one-zone benchmark model with $N_\mathrm{n} = 3.16\times 10^{13}\,\mathrm{cm}^{-3}$
for the CEMP-i star HE2148-1247 from Figure \ref{fig:denn3p16d13cycle} for comparison with the abundances in the star CS31062-050
we would find discrepancies between the predicted and observed abundances of Ba and Pr for the second star.
A comparison of the surface abundances of Ba and Pr predicted for the two values of $N_\mathrm{n}$ 
in Figures \ref{fig:denn3p16d14cycle} and \ref{fig:denn3p16d13cycle}
reveals that both elements have enhanced abundances in the case with the higher $N_\mathrm{n}$. Therefore,
the higher neutron density could eliminate the Ba abundance discrepancy for the star CS31062-050 in the case of the lower $N_\mathrm{n}$, 
and the assumption of a higher rate for the reaction
$^{141}$Ba(n,$\gamma)^{142}$Ba would significantly reduce the Pr abundance discrepancy at the same time.
The latter assumption would not affect the predicted Pr abundance for the star HE2148-1247 because at the lower neutron density,
that better matches its elemental abundance distribution in the selected range of $Z$, [Pr/Fe] has the strongest anti-correlation with a different
reaction, namely $^{141}$La(n,$\gamma)^{142}$La. 

\subsection{Multi-zone Monte Carlo simulation based on the RAWD model G}

\begin{figure}
  \centering
  \includegraphics[width=1.1\columnwidth]{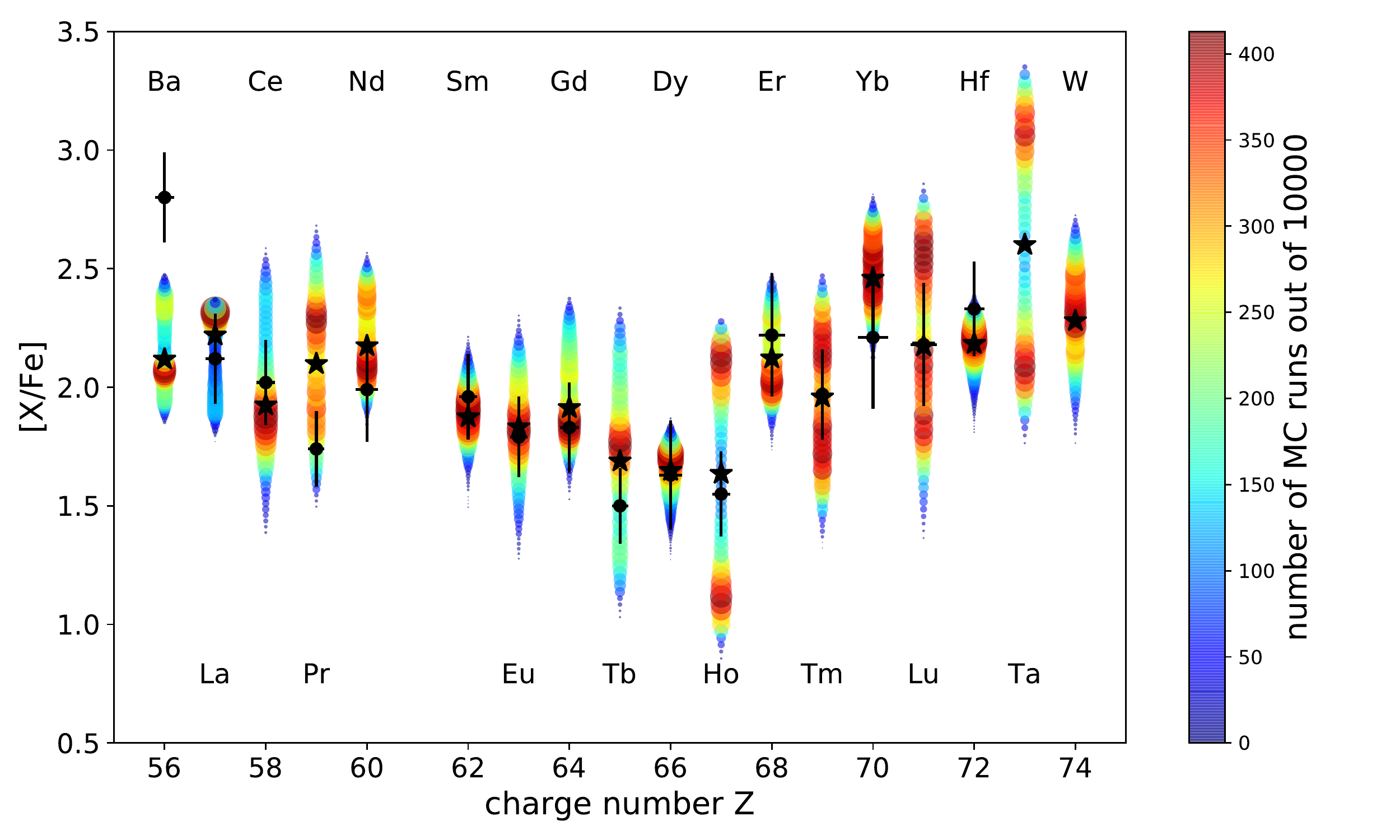}
  \caption{Same as in Figure \protect\ref{fig:uncert}, but for the multi-zone benchmark RAWD model G.}
  \label{fig:uncert3}
\end{figure}

Multi-zone benchmark models of the \ipr\ nucleosynthesis better simulate the physical conditions at which it is believed to occur in stars.
So far, only two such models have been investigated --- for the H ingestion by the He-shell convection in Sakurai's object \citep{paper:herwig-2011}
and in RAWDs \citep{denissenkov:17,denissenkov:19}. Here, we use the RAWD model G from the latter paper to simulate the \ipr\ nucleosynthesis of
the elements in the selected range of $Z$ in the He convective zone and to study its reaction rate uncertainties. In this model, neutrons are produced near the bottom of the He shell
in the reaction $^{13}$C($\alpha$,n)$^{16}$O for which fresh $^{13}$C is supplied by the positron decay of $^{13}$N that is made
in the reaction $^{12}$C(p,$\gamma)^{13}$N in the middle of the convective zone using protons entrained from the H-rich envelope.
The evolution of the maximum neutron density in the He zone of this model is shown in Figure \ref{fig:iRAWDNn} (the blue curve),
and the results of the MC simulation based on this multi-zone benchmark model are plotted in Figure \ref{fig:uncert3} and listed
in Table \ref{table:tab3}. By comparing Figures \ref{fig:uncert2} and \ref{fig:uncert3} and Tables \ref{table:tab2} and \ref{table:tab3},
we conclude that, in spite of its simplicity compared to the multi-zone model, the one-zone model with the constant neutron density
$N_\mathrm{n} = 3.16\times 10^{13}\,\mathrm{cm}^{-3}$ close to the peak value of $N_\mathrm{n,max}$ in the multi-zone model
can reliably predict both, distributions of the elemental abundances in the MC simulation and the identified key reactions
for most of the elements (the only exception is Sm, for Tb, Tm, Lu, and W the reactions with the largest and the second largest
correlation coefficient having swapped their places).
This means that one-zone MC simulations can be used instead of much more computationally time-consuming multi-zone MC simulations
for \ipr\ reaction rate uncertainty studies, provided that the former use neutron densities similar to the maximum ones found in the latter.

\begin{table}
        \caption{Same as Table \ref{table:tab1}, but for the one-zone benchmark model
           with $N_\mathrm{n}=3.16\times 10^{13}\,\mathrm{cm}^{-3}$.}
        \centering
        \begin{tabular}{|c|l|c|}
                \hline
                Element & Reaction & $r_\mathrm{P}(f_i,X_k/X_{k,0})$\\
                \hline
                Ba & {$^{134}$I} & +0.3689\\
                   & {$^{137}$Cs} & -0.6842\\
                \hline
                La & {$^{139}$Cs} & -0.2558\\
                   & {$^{139}$Ba} & -0.8651\\
                \hline
                Ce & {$^{139}$Ba} & +0.3907\\
                   & {$^{140}$Ba} & -0.7482\\
                \hline
                Pr & {$^{141}$Ba} & -0.4309\\
                   & {$^{141}$La} & -0.7339\\
                \hline
                Nd & {$^{143}$La} & -0.1513\\
                   & {$^{144}$Ce} & -0.8490\\
                \hline
                Sm & {$^{147}$Pr} & -0.6262\\
                   & {$^{149}$Nd} & -0.3377\\
                \hline
                Eu & {$^{151}$Nd} & -0.8124\\
                   & {$^{153}$Pm} & -0.2774\\
                \hline
                Gd & {$^{156}$Sm} & -0.5144\\
                \hline
                Tb & {$^{159}$Eu} & -0.8450\\
                   & {$^{159}$Gd} & -0.1872\\
                \hline
                Dy & {$^{162}$Gd} & -0.3019\\
                   & {$^{163}$Tb} & -0.6635\\
                \hline
                Ho & {$^{165}$Tb} & -0.2126\\
                   & {$^{165}$Dy} & -0.7440\\
                \hline
                Er & {$^{166}$Dy} & -0.8271\\
                \hline
                Tm & {$^{169}$Ho} & -0.6395\\
                   & {$^{169}$Er} & -0.5734\\
                \hline
                Yb & {$^{171}$Er} & -0.2703\\
                   & {$^{172}$Er} & -0.7914\\
                \hline
                Lu & {$^{175}$Tm} & -0.6339\\
                   & {$^{175}$Yb} & -0.5059\\
                \hline
                Hf & {$^{177}$Yb} & -0.3316\\
                   & {$^{178}$Yb} & -0.8157\\
                \hline
                Ta & {$^{181}$Lu} & -0.3271\\
                   & {$^{181}$Hf} & -0.7143\\
                \hline
                W  & {$^{182}$Hf} & -0.4016\\
                   & {$^{184}$Hf} & -0.6902\\
                \hline
        \end{tabular}
        \label{table:tab2}
\end{table}

\begin{table}
        \caption{Same as Table \ref{table:tab1}, but for the multi-zone benchmark RAWD model G.}
        \centering
        \begin{tabular}{|c|l|c|}
                \hline
                Element & Reaction & $r_\mathrm{P}(f_i,X_k/X_{k,0})$\\
                \hline
                Ba & {$^{137}$Cs} & -0.7871\\
                   & {$^{138}$Cs} & -0.2519\\
                \hline
                La & {$^{139}$Ba} & -0.9481\\
                \hline
                Ce & {$^{139}$Ba} & +0.5831\\
                   & {$^{140}$Ba} & -0.5864\\
                \hline
                Pr & {$^{141}$La} & -0.7828\\
                   & {$^{141}$Ce} & -0.3623\\
                \hline
                Nd & {$^{143}$Ce} & -0.1598\\
                   & {$^{144}$Ce} & -0.8230\\
                \hline
                Sm & {$^{147}$Nd} & -0.4748\\
                   & {$^{149}$Nd} & -0.4263\\
                \hline
                Eu & {$^{151}$Nd} & -0.5975\\
                   & {$^{151}$Pm} & -0.4975\\
                \hline
                Gd & {$^{155}$Sm} & +0.1868\\
                   & {$^{156}$Sm} & -0.6978\\
                \hline
                Tb & {$^{159}$Eu} & -0.5487\\
                   & {$^{159}$Gd} & -0.6169\\
                \hline
                Dy & {$^{162}$Tb} & -0.3479\\
                   & {$^{163}$Tb} & -0.5720\\
                \hline
                Ho & {$^{165}$Dy} & -0.8026\\
                \hline
                Er & {$^{135}$Xe} & -0.1977\\
                   & {$^{166}$Dy} & -0.7868\\
                \hline
                Tm & {$^{169}$Ho} & -0.2863\\
                   & {$^{169}$Er} & -0.8080\\
                \hline
                Yb & {$^{171}$Er} & -0.2324\\
                   & {$^{172}$Er} & -0.7638\\
                \hline
                Lu & {$^{175}$Tm} & -0.2777\\
                   & {$^{175}$Yb} & -0.7226\\
                \hline
                Hf & {$^{177}$Yb} & -0.3818\\
                   & {$^{178}$Yb} & -0.5322\\
                \hline
                Ta & {$^{181}$Hf} & -0.7721\\
                \hline
                W  & {$^{182}$Hf} & -0.6755\\
                   & {$^{184}$Hf} & -0.3709\\
                \hline
        \end{tabular}
        \label{table:tab3}
\end{table}

\section{Low-mass metal-poor AGB stars as an alternative site of i-process nucleosynthesis}
\label{sec:agb}   

Our paper is mainly focused on identifying the (n,$\gamma$) reaction rates whose uncertainties have
the strongest impact on the predicted abundances of the elements from Ba to W for the \ipr\ neutron densities 
and necessary neutron exposures that are encountered without any ad-hoc assumptions in RAWD stellar evolution  models 
and that agree with elemental abundances observed in the CEMP-i stars CS31062-050 and HE2148-1247. 
A potential alternative \ipr\ site are low-mass and low-Z AGB thermal pulse stars \citep{cristallo:16,Choplin:21}. 
The fact that H ingestion can take place in low-Z AGB stars is well established, and that \ipr\ conditions can be 
achieved is also suggested by several models \citep[for another example see][Sections 3.2.1 and 4.6]{Ritter:2018kb}. 
The question is if \iprn\ in low-Z AGB stars can produce abundance patterns observed in CEMP-i stars. 
\cite{karinkuzhi:20} concluded, based on comparing observations with models of \cite{Choplin:21}, that AGB stars
with $M = 1\,M_\odot$ and $-3\leq$\,[Fe/H]\,$\leq -2$ are sufficient in explaining \ipr\ abundances in CEMP-i stars and 
that RAWDs may not be needed for this.

The challenge of reproducing the \ipr\ abundance patterns in CEMP-i stars is that models need to satisfy two key 
observational properties of CEMP-i stars. First, they must reach a high enough neutron density to produce the ratios of abundances of
nearby elements at and just beyond the second peak, most notably [Ba/Eu]. Second, they must achieve high enough neutron exposures 
in order to reach the high hs/ls ratios of CEMP-i stars. As already noted by \cite{cristallo:16}, unless one makes 
ad-hoc assumptions on continued convective mixing, when standard stellar evolution predicts a split of the convection zone 
due to the energy input from convective-reactive $^{12}$C(p,$\gamma$) burning, the H ingestion and the simultaneous burning and mixing 
cannot be sustained for long enough to continue the convective-reactive generation of high neutron densities.

RAWD models solve both of these challenges without any ad-hoc
  mixing assumptions \citep{denissenkov:17,denissenkov:19}, and it is
  easy to see why. When the convection zone splits in low-Z AGB HIF
  simulations as a result of the H ingestion this will essentially end
  the mixing conditions needed for \iprn, because the
  $^{12}$C(p,$\gamma$) reaction that creates the neutron source
  $^{13}$C is separated from the hot conditions where the
  $^{13}$C($\alpha$,n) reaction releases neutrons fast enough to
  result in high neutron densities. Thus, when the split occurs (at
  least the way it occurs in standard 1D models) the \ipr\ conditions
  are quenched. The split happens when the energy released by
  the $^{12}$C(p,$\gamma$) reaction produces locally more entropy than
  can be carried away through convection driven by the underlying He
  burning. Then, at the location of the H burning inside the He-shell
  convection zone the entropy added by H burning will cause an entropy
  step as seen in stellar evolution models and in 3D hdyrodynamic
  simulations, such as Figure 3 in \cite{herwig:14}. The entropy step
  of course implies a radiative separation layer between the hot
  He-burning layer below and the cooler H-burning layer above,
  constituting the split that occurs roughly when $L_\mathrm{H}
  \ga L_\mathrm{He}$ \citep[for example][Figure 4]{paper:herwig-1999}. 
  Of course $L_\mathrm{H}$ depends on the entrainment rate of H
  which in turn depends on the He luminosity driving the convection
  \citep[see Figure 5 in][]{denissenkov:19}. RAWD thermal pulses have a
  key difference compared to AGB thermal pulses where the RAWD mass is
  the same as the AGB core mass. Due to their formation history the WD
  in a rapidly accreting system is cold and therefore the He-shell
  flashes are stronger. As a result, the entropy produced from burning
  of entrained H can continuously be distributed accross the
  convection zone due to the more efficient convection driven by
  higher $L_\mathrm{He}$, although the entrainment rate and therefore
  $L_\mathrm{H}$ is also somewhat larger. As a result, He-shell flashes
  in RAWDs have generally no split of the convection zone, and maybe
  just very weak events trailing the prolonged main H ingestion such
  as shown in Figure 3 of \cite{denissenkov:17}. This absence of a
  split during H-ingestion phases is a natural behaviour of RAWD
  models and does not require any forcing or ad-hoc assumptions and it
  leads to models with high neutron density \emph{and} high neutron
  exposure.

The formation of the split in the models used by \cite{karinkuzhi:20} is well documented 
in Figure 2 of \citet{Choplin:21}. As explained in the figure caption the time span between when 
the peak neutron density is reached and the time of the split is \unit{17.5}{\hour}. 
As a side we note that this is about or even less than the convective turn-over time of 
the He-shell flash convection zone in a \unit{1}{\Msun} low-Z AGB thermal pulse according 
to the NuGrid model database \citep[Figure 15 in][]{Ritter:2018kb}. However, in the stellar evolution models the energy transport and mixing
properties of a convective region are approximated in 1D with the mixing length theory
\citep[MLT,][]{cox:68} that is based on averaged quantities over space
and time. Certainly, time averages to describe a convective phenomenon that lasts just 
one convective turn-over become a poor approximation of the 3D hydrodynamic response 
to a dramatic energy input in a HIF that, when integrated over a convective turn-over time,  
typically reaches a substantial fraction of the binding energy of the convection 
zone \citep{jones:16,Ritter:2018kb,Clarkson:2020cq}. But if the stellar evolution simulation is 
to be taken seriously the split will generally end the \ipr\ nucleosynthesis 
without additional ad-hoc assumptions  \citep{cristallo:16}. This split will therefore limit 
the neutron exposure and along with it the hs/ls ratio that can be reached in low-Z AGB \ipr\ models.

\begin{figure}
  \centering
  \includegraphics[width=\columnwidth]{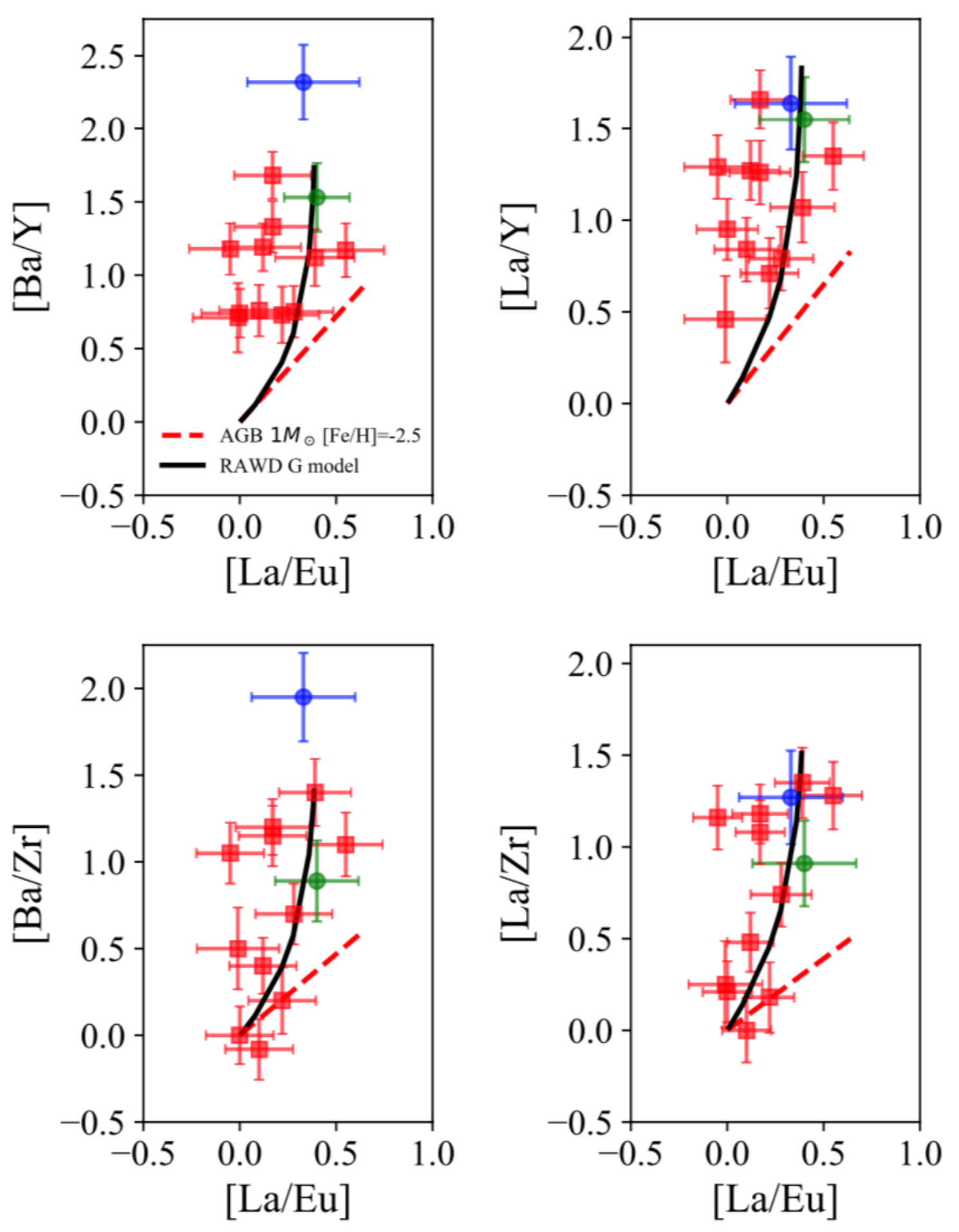}
  \caption{This plot is made to resemble Figure 13 of \protect\cite{karinkuzhi:20}, except that it shows only their CEMP-r/s stars and
           elemental abundance ratio dilution curve for the $1\,M_\odot$ AGB star model with [Fe/H]\,$=-2.5$ (dashed red curves). For a comparison, we have added
           the stars CS31062-050 (the blue circle) and HE2148-1247 (the green circle) and the dilution curve for our RAWD G model (solid black curves).}
  \label{fig:MyK}
\end{figure}

This deficiency is quite evident from the model to observations comparison in \cite{karinkuzhi:20}. 
Our Figure~\ref{fig:MyK} is made to resemble Figure 13 of \cite{karinkuzhi:20}, but we keep in it only their observed CEMP-r/s
  stars and abundance ratio dilution curves of the $1\,M_\odot$ AGB
  model with [Fe/H]\,$=-2.5$ because this is close to both the 
  metallicities of our RAWD F and G models, [Fe/H]\,$=-2.3$ and
  [Fe/H]\,$=-2.6$, respectively, and the mean [Fe/H] ratio of the
  CEMP-i stars.  For a comparison we have plotted the dilution
  curves of our RAWD G model (they are the same for the RAWD F model)
  and we have also added the stars CS31062-050 and HE2148-1247 studied
  in this paper.  The elemental abundance ratios plotted along the
  vertical axes measure the relative heights of the second- and
  first-peak regions, while the [La/Eu] ratio plotted along the
  horizontal axes is a standard diagnostic, similar to [Ba/Eu], used
  to classify a CEMP star either as CEMP-s, for [La/Eu]\,$> 0.5$,
  or CEMP-r/s, for $0<\mathrm{[La/Eu]} < 0.5$ \citep{beers:05}. 
This comparison of the low-Z AGB and RAWD models shows the effect of the split limiting the neutron exposure and 
thus the maximum hs/ls (such as Ba or La over Y or Zr) in the low-Z AGB model vs the much higher neutron exposure 
and hs/ls values in agreement with observations that can be reached in RAWD models. 

We also agree with \cite{Choplin:21} that their models have a too small dilution, i.e.\ a lot of 
low-Z AGB material would have to be accreted. 
From their equation (10), the ratio of the accreted mass with \ipr\ yields to the envelope mass of 
the accreting star is $M_\mathrm{acc}/M_\mathrm{env} = 1/f - 1$, where $f$ is the dilution coefficient. 
For their $f\approx 0.2$ $M_\mathrm{acc}/M_\mathrm{env}\approx 4$ which is impossible if $M_\mathrm{env}$ is the mass of 
the convective envelope of a CEMP-i red giant (11 out of 14 of their considered CEMP-i stars are red giants). 
For RAWD values of $f\approx 0.996$  $M_\mathrm{acc}/M_\mathrm{env}\approx 0.004$ 
which instead represents a more reasonable value. 

In view of these challenges the conclusion of \cite{karinkuzhi:20} that \iprn\ responsible for the abundance patterns 
observed in CEMP-i stars takes  place in low-mass, low-metalliticy AGB stars, and that other sites, such as the RAWDs, 
may not be needed is premature. This is in agreement with \cite{cristallo:16} who were the first to report heavy-element 
abundance distributions obtained as a result of
n-capture nucleosynthesis at the peak value of $N_\mathrm{n}\sim 10^{15}\ \mathrm{cm}^{-3}$ reached during the PIE in the He convective zone of 
a low-mass ($1.3 M_\odot$) low-$Z$ ([Fe/H]\,$=-2.85$) AGB model experiencing its first thermal pulse that, unlike
the AGB star model of \cite{karinkuzhi:20}, then followed the standard AGB evolution with further TPs, \spr\ nucleosynthesis,
and third dredge-ups. Just as demonstrated in our Figure~\ref{fig:MyK},
\cite{cristallo:16} concluded that the PIEs in AGB stars had unlikely been the source of the \ipr\ abundances
in CEMP-i stars because their models could not attain the high second- to first-peak n-capture elemental abundance ratios
observed in these stars.

We will close this section with an interesting twist to this argument which just demonstrates that the actual site of 
the \iprn\ for CEMP-i stars remains still not entirely settled. A similar situation to the question of whether low-Z AGB stars can or 
cannot account for CEMP-i star abundances exists for Sakurai's object. \cite{paper:herwig-2011} showed that 
post-processing simulations of the plain stellar evolution model of the very-late thermal pulse born-again star 
would not provide high enough neutron exposures to reach in this case the very low hs/ls. 
In Sakurai's object only the first-peak elements are enhanced. Sakurai's object is a nearby Pop I object. 
Since the initial Fe abundance is much higher than in CEMP-i stars a similar neutron exposure is expected to reach 
only to the first peak. However, due to the problem of the split, just as in low-Z AGB models, the presently favoured 
post-AGB very-late thermal pulse model, if post-processed as is, cannot reproduce the observed abundances.  
\cite{paper:herwig-2011} showed that if mixing accross the split was assumed to continue for \unit{1000}{\minute} 
then the observed abundances could be reproduced. 3D hydrodynamic simulations \citep{paper:herwig-2014} confirmed 
the early split predicted by 1D stellar evolution. It is interesting to note that RAWD models of high Z similar to 
that of Sakurai's object do an excellent job of explaining Sakurai's abundance pattern, just as the low-Z RAWD models 
do an excellent job in explaining CEMP-i stars. From a nucleosynthesis point of view  Sakurai's object is favoured 
to be a Pop~I RAWD. If Sakurai's object is instead a born-again AGB star then our 1D stellar evolution models are 
fundamentally unable to predict the occurrance and properties of the split correctly, and this would then also apply 
to the low-Z AGB stellar evolution models.

\section{Summary and conclusion}
\label{sec:summary}

For the neutron densities $N_\mathrm{n} = 3.16\times 10^{14}\,\mathrm{cm}^{-3}$ and
$N_\mathrm{n} = 3.16\times 10^{13}\,\mathrm{cm}^{-3}$ that were presumably attained in the \ipr\ stellar sites
that polluted the CEMP-i stars CS31062-050 and HE2148-1247, respectively,
we have identified the (n,$\gamma$) reaction rates of the 164 unstable isotopes selected for this study (Figure~\ref{fig:george})   
whose variations $f_i$ within the ranges constrained by
the Hauser-Feshbach model computations have the strongest impact on the predicted mass fractions
$X_k$ of 18 elements from Ba to W. This has been done by calculating the Pearson product-moment correlation coefficients
$r_\mathrm{P}(f_i,X_k/X_{k,0})$, where $X_{k,0}$ are the abundances predicted by the benchmark \ipr\
nucleosynthesis models for which all $f_i = 1$. Up to two
maximum coefficients with $|r_\mathrm{P}| > 0.15$ are presented in Tables \ref{table:tab1}, \ref{table:tab2}, and \ref{table:tab3}.
For each of the 18 selected elements, they identify one or two key reactions whose rate variations within their
theoretical uncertainty limits have the strongest impact on its predicted abundance and whose rates need therefore
to be measured experimentally first to improve the predictive power of stellar \ipr\ nucleosynthesis models. 
Among those, experimental measurements of the rates of the reactions $^{135}$I(n,$\gamma)^{136}$I, 
$^{137}$Cs(n,$\gamma)^{138}$Cs, $^{141}$Ba(n,$\gamma)^{142}$Ba,
and $^{141}$La(n,$\gamma)^{142}$La have the high priority because 
these rates have the strongest impact on the predicted abundances of Ba and Pr that have been observed in the considered CEMP-i
stars and these elements are particularly sensitive to neutron density.
Our MC simulations provide in addition statistically meaningful nuclear uncertainties for the \ipr\ abundance predictions. 
This enables a quantitative comparison with observations, e.g. compare the predicted and observed abundances of Ba and Pr in
Figure~\ref{fig:uncert2} where their error bars are shown. As Figure~\ref{fig:uncert} shows, with the currently large nuclear uncertainties 
our \ipr\ model predictions agree with observations, except for Pr, and, in the cases of low $N_\mathrm{n}$ (Figures~\ref{fig:uncert2} and \ref{fig:uncert3}) for Ba.

We have shown that the uncertainties of the temperature-dependent $\beta$-decay rates of the same 164 unstable isotopes have a much weaker effect
on the predicted abundances and that, at least, uncertainty studies of the elements from Ba to W can rely on one-zone MC simulations,
instead of computationally time-consuming multi-zone MC simulations, provided that the former use values of $N_\mathrm{n}$
consistent with those attained in the latter.

In order to underpin the astrophysical context of this impact study we have  discussed the question whether recently proposed 
low-mass low-Z AGB star models \citep{karinkuzhi:20,Choplin:21} could be
an alternative site of the \ipr\ nucleosynthsis for the heavy-element
abundance enhancements in CEMP-i stars. We have shown that the ratios of the second- to the first-peak n-capture elemental
abundances predicted by these AGB models are not as high as those observed in CEMP-i stars that, on the other hand,  are well reproduced
by our low-Z RAWD models. Therefore, in agreement with the results of \cite{cristallo:16},
we conclude that the low-mass low-Z AGB stars are not as likely candidates for the source of the \ipr\ abundances for
CEMP-i stars as the RAWDs. However, the treatment of convective-reactive H-ingestion flash events is 
not well approximated by 1D stellar evolution, and future 3D hydrodynamic simulations will have 
to reveal the true nature of the formation and evolution of the He-shell convection split and 
the subsequent convective mixing and nucleosynthesis.

The computational and analysis tools developed during our work on Papers I, II and this paper can be used to extend our analysis
to other neutron densities and to other reactions and elements.

\section*{Data availability}

Figures with correlations like those shown in Figure~\ref{fig:corr1} for all of the elements listed in Table~\ref{table:tab1}
can be found at \url{https://doi.org/10.5281/zenodo.4148667}. Other data underlying this article can be shared on reasonable 
request to the corresponding author.

\section*{Acknowledgements}

FH acknowledges funding from NSERC through a Discovery Grant. This
research is supported by the National Science Foundation (USA) under
award No. PHY-1430152 (JINA Center for the Evolution of the Elements).
The authors thank Iris Dillmann, Barry Davids, Chris Ruiz and Artemis Spyrou for fruitful
discussions of this problem.
The computations for this research were carried out on Compute Canada machines Niagara operated 
by SciNet and the Arbutus cloud operated by RCS at the University of Victoria. 
We appreciate the work of many researchers who have contributed to the development of
the NuGrid computer codes used in this study. 
We would also like to thank the anonymous referee whose comments have helped us to significantly
improve the paper. 




\bibliographystyle{mnras}
\bibliography{paper.bib}



\bsp	
\label{lastpage}
\end{document}